\documentclass[twocolumn]{openjournal}

\usepackage{lipsum}

\usepackage{xcolor}
\usepackage{textgreek}
\usepackage[utf8]{inputenc}
\usepackage[english]{babel}

\usepackage{hyperref}
\hypersetup{
    unicode, 
    colorlinks=true,
    linkcolor=linkcolor,
    citecolor=linkcolor,
    filecolor=linkcolor,
    urlcolor=linkcolor,
}
\usepackage{color,colortbl}
\definecolor{linkcolor}{rgb}{0.0,0.3,0.5}
\usepackage{tensind}
\tensordelimiter{?}
\DeclareGraphicsExtensions{.bmp,.png,.jpg,.pdf}
\usepackage{verbatim}
\usepackage[normalem]{ulem}
\usepackage{orcidlink}
\usepackage{soul}
\usepackage{amsmath,amssymb}
\usepackage{booktabs}
\usepackage{graphicx}
\usepackage{microtype}
\usepackage{xcolor}

\newcommand{\Msun}{M_\odot}

\graphicspath{ {./figs/} }
\makeatletter
\def\@hex@@Hex#1%
 {\if a#1A\else \if b#1B\else \if c#1C\else \if d#1D\else
  \if e#1E\else \if f#1F\else #1\fi\fi\fi\fi\fi\fi \@hex@Hex}
\makeatother
\definecolor{afcolor}{HTML}{b3443c}

\def\Msun{M_\odot}
\def\HH{$\rm H_{2}$~}
\def \fcold {f_{\rm{cold}}}
\def \fej {f^{\rm{ej}}_*}
\def \feff {f^{\rm{eff}}_*}
\def\msun{\Msun}

\graphicspath{ {./figs/} }

\begin{document}



\def\be{\begin{equation}}
\def\ee{\end{equation}}
\newcommand\code[1]{\textsc{\MakeLowercase{#1}}}
\newcommand\quotesingle[1]{`{#1}'}
\newcommand\quotes[1]{``{#1}"}
\def\gsim{\lower.5ex\hbox{\gtsima}} 
\def\lsim{\lower.5ex\hbox{\ltsima}} 
\def\gtsima{$\; \buildrel > \over \sim \;$} 
\def\ltsima{$\; \buildrel < \over \sim \;$} \def\gsim{\lower.5ex\hbox{\gtsima}} 
\def\lsim{\lower.5ex\hbox{\ltsima}} 
\def\simgt{\lower.5ex\hbox{\gtsima}} 
\def\simlt{\lower.5ex\hbox{\ltsima}}

\def\msun{{\rm M}_{\odot}}
\def\lsun{{\rm L}_{\odot}}
\def\dsun{{\cal D}_{\odot}}
\def\fsun{\xi_{\odot}}
\def\zsun{{\rm Z}_{\odot}}
\def\msunyr{\msun {\rm yr}^{-1}}
\def\gdens{\msun\,{\rm kpc}^{-2}}
\def\sfrdens{\msun\,{\rm yr}^{-1}\,{\rm kpc}^{-2}}

\def\mum{\mu {\rm m}}
\newcommand{\angstrom}{\mbox{\normalfont\AA}}
\def\cc{\rm cm^{-3}}
\def\uflux{{\rm erg}\,{\rm s}^{-1} {\rm cm}^{-2} }

\def\fdust{\xi_{d}}
\def\fesc{f_{\rm esc}}
\def\td{\tau_{sd}}
\def\Sg{$\Sigma_{g}$}
\def\S*{$\Sigma_{\rm SFR}$}
\def\Ssfr{\Sigma_{\rm SFR}}
\def\Sgas{\Sigma_{\rm g}}
\def\Sstar{\Sigma_{\rm *}}
\def\Sesc{\Sigma_{\rm esc}}
\def\Srad{\Sigma_{\rm rad}}
\def\sSFR{\rm sSFR}

\def\Dsolar{${\cal D}/\dsun$}
\def\Zsolar{$Z/\zsun$}
\def\DDsolar{\left( {{\cal D}\over \dsun} \right)}
\def\ZZsolar{\left( {Z \over \zsun} \right)}
\def\kms{{\rm km\,s}^{-1}\,}
\def\skms{$\sigma_{\rm kms}\,$}

\def\Scii{$\Sigma_{\rm [CII]}$}
\def\Sciimax{$\Sigma_{\rm [CII]}^{\rm max}$}
\def\CII{\hbox{[C~$\scriptstyle\rm II $]~}}
\def\CIII{\hbox{C~$\scriptstyle\rm III $]~}}
\def\OII{\hbox{[O~$\scriptstyle\rm II $]~}}
\def\OIII{\hbox{[O~$\scriptstyle\rm III $]~}}
\def\HH{\hbox{H$_2$}~} 
\def\HI{\hbox{H~$\scriptstyle\rm I\ $}} 
\def\HII{\hbox{H~$\scriptstyle\rm II\ $}} 
\def\CIion{\hbox{C~$\scriptstyle\rm I $~}}
\def\CIIion{\hbox{C~$\scriptstyle\rm II $~}}
\def\CIIIion{\hbox{C~$\scriptstyle\rm III $~}}
\def\CIVion{\hbox{C~$\scriptstyle\rm IV $~}}
\def\nhh{n_{\rm H2}}
\def\nhi{n_{\rm HI}}
\def\nhii{n_{\rm HII}}
\def\fhh{x_{\rm H2}}
\def\fhi{x_{\rm HI}}
\def\fhii{x_{\rm HII}}
\def\fd{f^*_{\rm diss}} 
\def\ks{\kappa_{\rm s}}

\def\cyan{\color{cyan}}
\definecolor{apcolor}{HTML}{b3003b}
\definecolor{afcolor}{HTML}{800080}
\definecolor{lvcolor}{HTML}{DF7401}
\definecolor{mdcolor}{HTML}{01abdf} 
\definecolor{cbcolor}{HTML}{ff0000}
\definecolor{sccolor}{HTML}{cc5500} 
\definecolor{sgcolor}{HTML}{00cc7a}

\title{Top-heavy Initial Mass Functions are not enough at cosmic dawn}

\author{Andrea Ferrara$^{1,2}$\orcidlink{0000-0002-9400-7312}}
\email{andrea.ferrara@sns.it}
\affiliation{$^1$Scuola Normale Superiore,  Piazza dei Cavalieri 7, Pisa I-56126, Italy}
\affiliation{$^2$Centro B. Segre, Accademia Nazionale dei Lincei, Via della Lungara, 10, Roma I-00165, Italy}

\author{Pratika Dayal$^{3,4,5}$\orcidlink{0000-0001-8460-1564}}
\email{pdayal@cita.utoronto.ca}
\affiliation{$^3$Canadian Institute for Theoretical Astrophysics, 60 St George St, University of Toronto, Toronto, ON M5S 3H8, Canada}
\affiliation{$^4$David A. Dunlap Department of Astronomy and Astrophysics, University of Toronto, 50 St George St, Toronto ON M5S 3H4, Canada}
\affiliation{$^5$Department of Physics, 60 St George St, University of Toronto, Toronto, ON M5S 3H8, Canada}

\begin{abstract}
A top-heavy stellar initial mass function (IMF) has been proposed to explain the unexpected abundance of ultraviolet (UV)-bright galaxies at cosmic dawn. We test this
possibility using the metallicity-dependent, top-heavy IMF recently proposed by \citet{Chon24}, which is motivated by simulations of metal-poor star formation. We compare this IMF with a Salpeter IMF, self-consistently including the associated changes in UV emission, supernova feedback, metal enrichment, and dust production. At fixed star-formation rate, the Chon IMF enhances the intrinsic UV luminosity by a factor
${\cal B}_{1500}\simeq4.7$--$2.8$ over halo masses
$M_h=10^8$--$10^{12}\,\Msun$. However, its larger supernova rate suppresses
star formation and increases dust attenuation, reducing the net boost to
${\cal B}_{\rm UV}\simeq2.0$--$1.06$. For halos relevant to the bright end of
the UV luminosity function, the enhancement is only
${\cal B}_{\rm UV}\simeq1.2$--$1.4$, or about $0.3$ mag. Calculations with
the {\sc delphi} semi-analytic model confirm that the self-consistent Chon and
Salpeter UV luminosity functions differ by less than $0.5$ dex at
$z\simeq11$--$17$. Thus, a top-heavy IMF alone cannot explain the
bright galaxy excess. Conversely, the predicted pair-instability supernova
abundance differs by factors of $2$--$6$ between the two IMFs, providing a
potentially cleaner observational diagnostic of the high-mass stellar
populations at these early epochs.
\end{abstract}

\maketitle

\section{Introduction}
The discovery of unexpectedly abundant, UV-bright galaxies in the first few
hundred Myr after the Big Bang (e.g., \citealt{Naidu22, Harikane22, Castellano23, Bunker23, Finkelstein24, Curtis23, Carniani24a,  Naidu25}) has reopened a basic question in galaxy
formation: are these sources bright because galaxies formed stars more efficiently, because their dust attenuation was lower, or because their stellar populations produced more UV light per unit star formation? The last possibility has attracted some attention after the launch of JWST, since a stellar initial mass function (IMF) biased toward massive stars would naturally increase both the non-ionizing UV luminosity and the production rate of ionizing photons. {We refer interested readers to a recent mini-review by \citet{fontana2026}}.

The IMF enters almost every inference made for high-redshift galaxies. It sets the UV luminosity per unit star formation rate, the ionizing photon production efficiency, the supernova (SN) rate, the metal and dust yields, and the stellar mass-to-light ratio (for reviews see \citealt{sommerville:2015, Dayal18}).  A top-heavy IMF is therefore an appealing solution to the apparent overabundance of bright galaxies at cosmic dawn: without changing the halo population, one can make galaxies brighter by placing a larger fraction of the newly formed stellar mass in short-lived,
massive stars.  In its simplest form, this argument amounts to increasing the conversion factor between SFR and UV luminosity, {or the light-to-mass ratio}.

Several recent studies have explored this possibility in the post-JWST context.  Spectroscopic arguments have been made from individual sources, most notably GS9422, whose nebular-continuum-dominated spectrum has been interpreted as evidence for an extremely top-heavy IMF capable of producing an ionizing emissivity far above that of standard stellar populations
\citep{Cameron24, Katz23, Katz25}.  However, the same object also illustrates the degeneracy of IMF inferences: alternative fits combining metal-poor stars with a weak active galactic nucleus (AGN) can reproduce much of the required ionizing spectrum without invoking such an extreme IMF \citep{Li24}.  

A second class of studies has tested whether IMF variations can affect the UV luminosity function itself. Galaxy-formation models have reached mixed conclusions. \citet{mauerhofer2025} have incorporated interstellar medium (ISM) physics from the {\sc SPHINX} simulations into the {\sc delphi} semi-analytic model. They track the assembly of each system and assign an IMF motivated by the simulations of \citet{Chon22} which becomes increasingly top-heavy with decreasing metallicity and increasing redshift. It is crucial to note that whilst resulting in an increase in both the UV light and ionizing photons produced per unit stellar mass, an increasingly top-heavy IMF strengthens supernova feedback and alters metal and dust production. {These authors show that matching to the observed UV LF at $z \geq 11$ require a decrease in dust attenuation, even when increasingly top-heavy IMFs are implemented.} Similar results have been found using the {\sc astraeus} model by \citet{Cueto23}, who also highlight the fact that matching to the $z \leq 11$ UV LF requires a decrease in the star formation efficiency in the evolving IMF model as compared to one that uses a Salpeter IMF. These self-consistent treatments lead to an important conclusion: an evolving IMF alone is unlikely to solve the abundance problem, because the luminosity gain is partly compensated by the {dust and} feedback penalties. 

Other implementations have found more optimistic outcomes when the IMF is linked to the most intensely star-forming systems.  For example, models in which the IMF becomes top-heavy above a density or star-formation threshold can reproduce the observed UV luminosity functions at \(z\simeq5-15\) \citep{Hutter_2024}.  Yet even in these models the UV enhancement depends strongly on whether galaxies are feedback regulated.  In low-mass systems, where supernova feedback controls the star-formation efficiency, the boost is limited; in more massive halos the same IMF variation can produce a larger change in \(M_{\rm UV}\).  This already suggests that the effect of a top-heavy IMF cannot be assessed from UV LF alone.

Finally, \citet{Trinca24} used the Cosmic Archaeology Tool to follow Pop~III/II stars and accreting black holes over \(4\leq z\leq20\).  In their reference model, stars dominate the UV luminosity function and AGN emission is subdominant at \(z\gtrsim10\), but the predicted
number density of \(M_{\rm UV}\simeq -20\) galaxies falls below some early JWST estimates at \(z\gtrsim12\).  They showed that a gradual IMF transition, modulated by metallicity and redshift and motivated by simulations  \citep{Chon21,Chon22}, increases the UV output per unit SFR and reduces the tension with the observed bright-end counts.

The same point emerges from cloud-scale radiation-hydrodynamic calculations.
A top-heavy IMF increases the UV luminosity-to-mass ratio, but it also
increases radiative feedback.  Depending on the dust-to-gas ratio and cloud
surface density, this can lower the star-formation efficiency and partially
erase the luminosity advantage \citep{Menon24}.  Thus the IMF and the star formation efficiency (SFE)
are physically coupled: making the stellar population more luminous also makes
it more capable of disrupting the gas from which it forms.

Cosmological simulations with variable IMFs reach a similar conclusion from a
different direction.  Density-dependent IMF models can brighten high-redshift
galaxies by factors of a few, but self-consistent metal and dust yields
introduce a delayed counter-effect: the same massive stars that increase the
far-UV luminosity also increase supernova feedback, metal enrichment, and dust
production \citep{Durrant26}.  Semi-analytic cloud models likewise show that
the IMF can strongly affect the UV luminosity function, but only as part of a
larger system involving cloud masses, burstiness, feedback timescales, and
star-formation efficiencies \citep{Cueto26}.

These studies motivate a more precise question: if the IMF is allowed to vary,
what key physical variable(s) should control the variation?  Several possibilities
have been proposed.  In primordial systems, the IMF is expected to be
top-heavy because molecular cooling, weak metal-line cooling, and low dust
abundances keep the characteristic fragmentation mass high
\citep{Omukai98,Bromm01,Schneider02, Omukai05,Hirano14,Hirano15, Hirano26}.  At later
times, metallicity becomes central because metals and dust open new cooling
channels, lowering the fragmentation mass and driving the IMF toward the
present-day form.  At high redshift, however, the CMB temperature floor also
matters: even metal-enriched gas cannot cool below
\(T_{\rm CMB}=2.73(1+z)\,{\rm K}\), so the minimum gas temperature, and hence
the characteristic stellar mass, may remain elevated
\citep{Schneider10,Chon22}.

This gives a physically motivated route to a variable IMF in which the degree
of top-heaviness is controlled primarily by metallicity and redshift.  In the
simulations of \citet{Chon21} and \citet{Chon22}, low-metallicity clouds form stellar
populations with enhanced massive-star fractions because the gas remains
warmer and fragments less efficiently.  The transition is not a discontinuous
Pop~III/Pop~II switch, but a gradual change in the stellar mass distribution
as enrichment proceeds and the CMB temperature changes.  This is the same
physical idea adopted phenomenologically by a number of works \citep{Cueto23, Trinca24, mauerhofer2025}, who describe
the UV effect using a composite Kroupa plus log-flat IMF whose relative weight
depends on \(z\) and \(Z\).

Other prescriptions emphasize different variables.  Empirical and
theoretical studies have connected IMF variations to the density, pressure,
temperature, and turbulent state of star-forming gas, motivated by the
expectation that the characteristic stellar mass should track the thermal or
turbulent Jeans mass \citep[e.g.][]{Padoan97,Elmegreen04,Marks12,
Narayanan12,Hennebelle13, Mathew26}.  In the integrated-galactic IMF framework, the
galaxy-wide IMF can also depend on the SFR, because galaxies with larger SFRs
sample more massive embedded clusters and hence produce a larger fraction of
massive stars \citep{Weidner11,Jerabkova18}.  These choices are useful for
galaxy-scale modeling, but they also introduce a potential ambiguity: the IMF
is then linked to quantities, such as SFR, sSFR, pressure, or gas density, that
are themselves affected by feedback from the same massive stars.  In contrast,
metallicity provides a direct connection to gas thermodynamics, cooling, and
fragmentation, while still being determined self-consistently by the evolving
stellar population.

For this reason we adopt the \citet[][]{Chon24} IMF as our fiducial example of a physically
motivated top-heavy IMF.  We adopt this IMF because it encodes a physically sound thermal fragmentation mechanism:
low metallicity increases the
characteristic stellar mass and flattens the high-mass IMF.  Moreover, the Chon et al. 
prescription gives an explicit metallicity dependence that can be cleanly inserted
into a closed loop: the metallicity determines the IMF; the IMF
determines the supernova rate, metal yield, return fraction, UV production,
and dust yield; these in turn determine the star-formation efficiency and the
next value of the metallicity.

In this paper we revisit the IMF explanation of the bright-end UV luminosity
function from this perspective.  We ask a deliberately restricted question:
how much does the emergent UV luminosity change when a fixed Salpeter IMF is
replaced by a metallicity-dependent Chon IMF, while the feedback, enrichment,
and dust terms are updated consistently? 

The paper is organized as follows\footnote{Throughout the paper, we assume a flat Universe with the following cosmological parameters: $\Omega_m = 0.3075$, $\Omega_{\Lambda} = 1- \Omega_{\rm M}$, and $\Omega_{b} = 0.0486$,  $h=0.6774$, $\sigma_8=0.826$, where $\Omega_{m}$, $\Omega_{\Lambda}$, and $\Omega_{b}$ are the total matter, vacuum, and baryon densities, in units of the critical density; $h$ is the Hubble constant in units of $100\,\kms \rm Mpc^{-1}$, and $\sigma_8$ is the late-time fluctuation amplitude parameter \citep{planck:2015}.}.  In Section~2 we define the Salpeter and
Chon IMF models and derive the IMF-dependent feedback, enrichment, and dust
quantities.  In Section~3 we compute the UV boost factors and show why the net
boost is much smaller than the direct stellar-population term.  In
Section~4 we implement the two IMF choices into the {\sc DELPHI} semi-analytic model and compare the resulting
UV luminosity functions.  {In Section~5 we compute the predicted
pair-instability supernova (PISN) rates and show that they provide a more sensitive
diagnostic of the IMF}.  We summarize our conclusions in Section~6.

\section{Baseline calculation}
At fixed halo mass $M_h$ and redshift $z$, we write the emergent UV
luminosity at 1500 \AA\ as
\begin{equation}\label{eq:L1500}
 L_{\rm UV}={\rm SFR}\,{\cal K}_{1500}\,T_d,
\end{equation}
where ${\cal K}_{1500}$ is the intrinsic 1500~\AA\ luminosity per unit star
formation rate and $T_d$ is the dust transmission.  To quantitatively assess the impact of IMF on the observed galaxy UV luminosity we compare a
metallicity-dependent Chon IMF \citep{Chon24} to a fixed Salpeter \citep{Salpeter55} reference IMF over the same
stellar mass interval, $0.1\leq m/\Msun\leq300$.  

\subsection{Reference and top-heavy IMFs}
The reference model is a fixed Salpeter IMF,
$\phi_{\rm Salp}(m)=A_{\rm Salp}\,m^{-2.35}$, defined\footnote{Note that, unlike most previous works, the adopted IMFs include Very Massive Stars (VMS) with $m=120$--$300\ M_\odot$} over
$0.1<m/\Msun<300$ and normalized by
\begin{equation}\label{eq:norm}
 \int_{0.1}^{300}m\phi_{\rm Salp}(m)\,dm=1.
\end{equation}
Thus $\phi(m)dm$ gives the number of stars in the interval $[m,m+dm]$
per unit stellar mass formed.
\begin{center}
    \begin{figure}[!t]
        \centering
    	\includegraphics[scale=0.42]{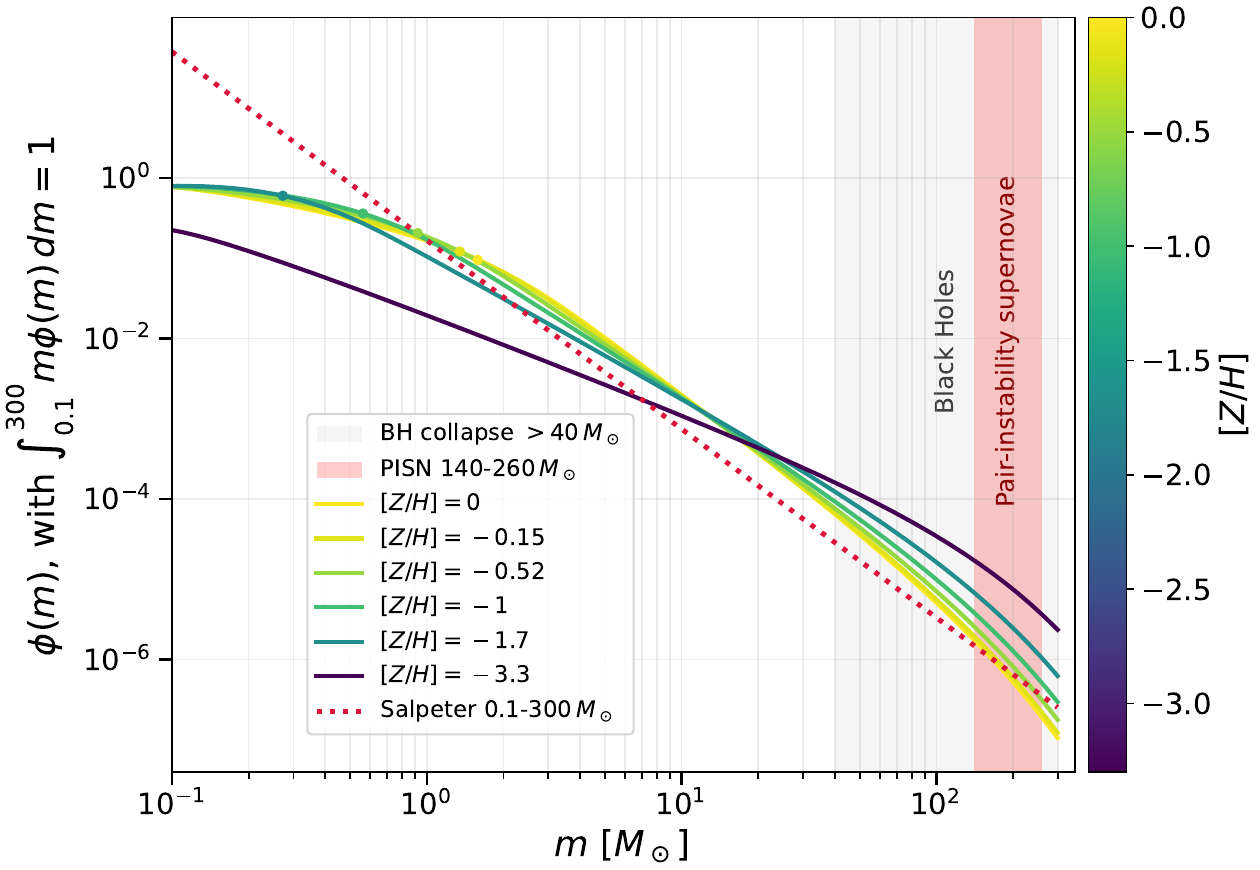}
    	\caption{Comparison between the metallicity-dependent Chon IMF and the fixed Salpeter
reference IMF.  All IMFs are evaluated and normalized over the same stellar
mass interval, $0.1<m/\Msun<300$. The Chon IMF is obtained from eq. \ref{eq:Chon} with parameters given therein. 
The colored curves show the Chon IMF at the six metallicities used in the
calculation, while the dotted red curve shows the Salpeter
$0.1$--$300\,\Msun$ reference IMF.  The colored points mark the characteristic
mass $m_0(Z)$ on each Chon IMF.  The grey shaded region indicates the mass
range $>40\,\Msun$ assumed to collapse directly into black holes, while the red shaded band marks the
$140$--$260\,\Msun$ pair-instability supernova range.}
        \label{fig:fig01}
    \end{figure}
\end{center}

We adopt the metallicity-dependent IMF proposed by \citet{Chon24} as a physically motivated example of a top-heavy IMF because it is not imposed phenomenologically, but arises from simulations of low-metallicity star formation in which reduced cooling and fragmentation naturally shift the characteristic stellar mass upward while preserving an explicit dependence on the gas metallicity. We note that no explicit dependence on redshift is included in the IMF.  

The IMF depends on gas metallicity, $Z$ and is written as
\begin{equation}\label{eq:Chon}
 \phi_{\rm Chon}(m|Z)
 =
 A_{\rm Chon}(Z)\,
 m^{-\alpha(Z)}\,
 \gamma(m,Z),
\end{equation}
with
\begin{equation}
 \alpha(Z)=2.33+0.33\log_{10}(Z/Z_\odot),
\end{equation}
\begin{align}
 \gamma(m,Z)
 &=
 \left[
 1-\exp\left(
 -\left[\frac{m}{m_0(Z)}\right]^c
 \right)
 \right]
 \nonumber\\
 &\quad\times
 \exp\left(
 -\frac{m_{\rm min}}{m}
 -\frac{m}{m_{\rm max}}
 \right),
\end{align}
and
\begin{equation}
 m_0(Z)=10^{0.2+0.45\log_{10}(Z/Z_\odot)}\,\Msun.
\end{equation}
Following the authors, we adopt
$m_{\rm min}=0.04\,\Msun, m_{\rm max}=150\,\Msun, c=1.6$. Note that $m_{\rm min}$ and $m_{\rm max}$ are intrinsic parameters of the Chon functional form, not hard integration limits.  As for the Salpeter IMF, the Chon IMF is evaluated over $0.1<m/\Msun<300$ normalized to unity as in eq. \ref{eq:norm}. 

\subsection{Supernova rate, metal yield, and return fraction}

For either IMF, stars in the interval $8<m/\Msun<40$ explode as
core-collapse supernovae (CCSN), while stars above $40\,\Msun$ collapse
directly into black holes.  Pair-instability supernovae in the interval $140$--$260\,\Msun$ are not included in the baseline feedback, metal, or dust budgets.  Including this channel could substantially modify metal and dust production at the lowest metallicities and should be treated as
a separate extension. However, as we will see later, they will be used as a passive tracer discriminating between a standard Salpeter and a top-heavy IMF, exemplified here by the Chon IMF. 

For an IMF $\phi_I$, where $I\in\{\mathrm{Chon},\mathrm{Salp}\}$, the CCSN
rate per unit stellar mass is then
\begin{equation}
 \nu_{{\rm SN},I}(Z_I)=
 \frac{\displaystyle\int_{8}^{40}\phi_I(m|Z_I)\,dm}
 {\displaystyle\int_{0.1}^{300}m\phi_I(m|Z_I)\,dm}.
\end{equation}
For the Salpeter reference, $\phi_{\rm Salp}(m|Z_{\rm Salp})$ simply reduces
to $\phi_{\rm Salp}(m)$. For a CCSN progenitor of mass $m$, the metal ejecta are \citep{Kim14, Ferrara25b}
\begin{align}
 M_Z(m)&=2.09M_{\rm O}(m)+1.06M_{\rm Fe}(m),\\
 M_{\rm O}(m)&=0.375\exp(-17.94/m)\,\Msun,\\
 M_{\rm Fe}(m)&=27.66\exp(-51.81/m)\,\Msun.
\end{align}
The fit is in very good agreement with the detailed results reported in \citet{Nomoto13}.
The mean metal yield per exploding CCSN is
\begin{equation}
 y_{Z,I}(Z_I)=
 \frac{\displaystyle\int_{8}^{40}M_Z(m)\phi_I(m|Z_I)\,dm}
 {\displaystyle\int_{8}^{40}\phi_I(m|Z_I)\,dm}.
\end{equation}

In the instantaneous-recycling approximation, stars below
$m_{\rm ret}=2.3\,\Msun$ live indefinitely.  Following \citet{Weidemann00}, we take the remnant mass as follows:
\begin{equation}
 w_m=\begin{cases}
 0.444+0.084m, & 2.3<m/\Msun<8,\\
 1.4\,\Msun, & 8<m/\Msun<40,\\
 m, & 40<m/\Msun<300.
 \end{cases}
\end{equation}
The return fraction is
\begin{equation}
 R_I(Z_I)=
 \frac{\displaystyle\int_{2.3}^{300}[m-w_m]\phi_I(m|Z_I)\,dm}
 {\displaystyle\int_{0.1}^{300}m\phi_I(m|Z_I)\,dm}.
\end{equation}
For later use, it is convenient to define
\begin{equation}
 P_{Z,I}(Z_I)\equiv
 y_{Z,I}(Z_I)\nu_{{\rm SN},I}(Z_I)[1-R_I(Z_I)].
\end{equation}

\subsection{Self-consistent metallicities}

For each IMF model $I\in\{{\rm Chon},{\rm Salp}\}$, the characteristic feedback velocity
associated with CCSNe is
\begin{equation}
 v_{s,I}^2(Z_I)=\frac{\nu_{{\rm SN},I}(Z_I)E_{\rm SN}}{\Msun},
\end{equation}
where $E_{\rm SN}=10^{51}\,{\rm erg}$ is the SN explosion energy, and $\Msun$ in the denominator is converted to grams.  Following \citet{dayal2014, Ferrara23, Ferrara26}  the star-formation efficiency is
\begin{equation}
 \epsilon_{\star,I}(M_h,Z_I)=
 \epsilon_0
 \frac{v_c^2(M_h,z)}
 {v_c^2(M_h,z)+f_wv_{s,I}^2(Z_I)},
\end{equation}
where $\epsilon_0=0.09,f_w=0.12$, and $v_c(M_h,z)=[10GH(z)M_h]^{1/3}$ is the halo circular velocity, where $H(z)$ is the Hubble constant.
These parameters give the best match to super-early galaxy properties and UV LF in the Attenuation-Free Model \citep[AFM,][]{Ferrara26}.

To compute the gas metallicity we adopt the derivation in \citet[][eq. 11]{Ferrara25b}:
\begin{equation}
\label{eq:metal}
 Z_I(M_h)=
 \frac{
 P_{Z,I}(Z_I)
 }{
 a[z,\epsilon_{\star,I}(M_h,Z_I)]
 },
\end{equation}
where
\begin{equation}
 a(z,\epsilon_\star)=6.23[-a_0-a_1(1+z)]
 \left(\frac{0.01}{\epsilon_\star}\right),
\end{equation}
with $a_0=0.25$, $a_1=-0.75$. This equation applies to both IMFs.  For the fixed Salpeter reference,
$\phi_{\rm Salp}$ does not depend on metallicity; therefore
$y_{Z,{\rm Salp}}$, $\nu_{{\rm SN},{\rm Salp}}$, $R_{\rm Salp}$, and
$v_{s,{\rm Salp}}$ are constants, and the equation is explicit once $M_h$
and $z$ are specified.  For the Chon IMF, instead, the IMF depends on
$Z_{\rm Chon}$ through $\alpha(Z_{\rm Chon})$ and $m_0(Z_{\rm Chon})$.
Consequently, the set 
$\{ y_{Z,{\rm Chon}},  \nu_{{\rm SN},{\rm Chon}},  R_{\rm Chon},  v_{s,{\rm Chon}}, 
 \epsilon_{\star,{\rm Chon}}\}$
must be updated self-consistently as the metallicity changes.  We solve this
scalar fixed-point problem numerically at each halo mass, obtaining
$Z_{\rm Chon}(M_h)$ and $Z_{\rm Salp}(M_h)$. The final values of the various
relevant quantities are reported in Table \ref{tab:imf_quantities} for the two IMFs.

\begin{table*}[t]
\centering
\caption{IMF-dependent quantities entering the enrichment, feedback, and dust
terms.  Both the fixed Salpeter reference IMF and the metallicity-dependent
Chon IMF are evaluated and normalized over $0.1<m/\Msun<300$.  The CCSN interval is $8<m/\Msun<40$; stars above
$40\,\Msun$ are assumed to collapse directly into black holes. PISN occur in the interval $140<m/\Msun<260$ and only for metallicities $Z < 0.1\ Z_{\odot}$; this channel is
reported as a diagnostic but is not included in the baseline feedback, metal, or
dust budgets.  The final column gives $\xi_d=y_d\nu_{\rm SN}$, with
$y_d=0.1\,\Msun$ per CCSN.}
\label{tab:imf_quantities}
\small
\begin{tabular}{llccccc}
\hline
IMF & $Z/Z_\odot$ &
$y_Z$ &
$R$ &
$\nu_{\rm SN}$ &
\(\nu_{\rm PISN}\) &
$\xi_d$ \\
 &  &
$(\Msun\,{\rm SN}^{-1})$ &
 &
$(\Msun^{-1})$ &
$(\Msun^{-1})$ &
\\
\hline
Chon & $5.0\times10^{-4}$ & 2.372 & 0.308 & $1.445\times10^{-2}$ & $1.010\times10^{-3}$ & $1.445\times10^{-3}$ \\
Chon & $2.0\times10^{-2}$ & 1.866 & 0.424 & $1.826\times10^{-2}$ & $3.507\times10^{-4}$ & $1.826\times10^{-3}$ \\
Chon & $1.0\times10^{-1}$ & 1.672 & 0.435 & $1.736\times10^{-2}$ & -- & $1.736\times10^{-3}$ \\
Chon & $3.0\times10^{-1}$ & 1.549 & 0.451 & $1.684\times10^{-2}$ & -- & $1.684\times10^{-3}$ \\
Chon & $7.0\times10^{-1}$ & 1.461 & 0.476 & $1.690\times10^{-2}$ & -- & $1.690\times10^{-3}$ \\
Chon & $1.0$                 & 1.425 & 0.490 & $1.710\times10^{-2}$ & -- & $1.710\times10^{-3}$ \\
\hline
Salpeter & $5.0\times10^{-4}$ & 1.495 & 0.190 & $6.595\times10^{-3}$ & $8.847\times10^{-5}$ & $6.595\times10^{-4}$ \\
Salpeter & $2.0\times10^{-2}$ & 1.495 & 0.190 & $6.595\times10^{-3}$ & $8.847\times10^{-5}$ & $6.595\times10^{-4}$ \\
Salpeter & $1.0\times10^{-1}$ & 1.495 & 0.190 & $6.595\times10^{-3}$ & -- & $6.595\times10^{-4}$ \\
Salpeter & $3.0\times10^{-1}$ & 1.495 & 0.190 & $6.595\times10^{-3}$ & -- & $6.595\times10^{-4}$ \\
Salpeter & $7.0\times10^{-1}$ & 1.495 & 0.190 & $6.595\times10^{-3}$ & -- & $6.595\times10^{-4}$ \\
Salpeter & $1.0$                 & 1.495 & 0.190 & $6.595\times10^{-3}$ & -- & $6.595\times10^{-4}$ \\
\hline
\end{tabular}
\vspace{0.2cm}
\end{table*}

\subsection{The three boost factors}
Armed with the above results, we now return to the analysis of eq. \ref{eq:L1500} with the aim of determining the amplitude of the relative UV luminosity variation associated with the two IMFs considered. For this purpose, for each of the three factors in eq. \ref{eq:L1500} we introduce the relative ratio, or \textit{boost factor}, ${\cal B}$, between the Chon and the reference Salpeter IMF. The net UV boost is then
\begin{equation}
 {\cal B}_{\rm UV}
 =
 {\cal B}_{\rm SFR}\,
 {\cal B}_{1500}\,
 {\cal B}_{\rm dust}.
\end{equation}

\subsubsection{Direct 1500~\AA\ boost}
The fiducial UV calibration is computed for a constant star-formation history
of duration and evaluated at a stellar age of $T=30\,{\rm Myr}$,  which is consistent with the typical inferred age of stars in super-early galaxies. 

For each IMF and each of the six metallicities, we use the SB99
instantaneous-burst spectra only as simple stellar population kernels.  If
$L_{1500,I}^{\rm SSP}(t,Z)$ is the 1500~\AA\ luminosity of an instantaneous
burst of mass $M_{\rm burst}=10^6\,\Msun$, including stellar and nebular
emission, then the luminosity per unit burst mass is
\begin{equation}
 \ell_{1500,I}^{\rm SSP}(t,Z)
 =
 \frac{L_{1500,I}^{\rm SSP}(t,Z)}{M_{\rm burst}}.
\end{equation}
For a constant SFR, the intrinsic 1500~\AA\ luminosity per unit star formation
rate is therefore
\begin{equation}
 {\cal K}_{1500,I}(T,Z)
 =
 \int_0^T
 \ell_{1500,I}^{\rm SSP}(t,Z)\,dt,
\end{equation}
where $dt$ is expressed in years, so that
${\cal K}_{1500}$ has units
${\rm erg\,s^{-1}\,\AA^{-1}}\,(\Msun\,{\rm yr^{-1}})^{-1}$.

We compute one ${\cal K}_{1500}$ grid for the fixed Salpeter IMF and one for
the Chon IMF.  The grids are interpolated linearly in
$\log(Z/Z_\odot)$ and $\log{\cal K}_{1500}$.  The direct 1500~\AA\ boost is
\begin{equation}
 {\cal B}_{1500}(M_h)
 =
 \frac{
 {\cal K}_{1500,{\rm Chon}}[T,Z_{\rm Chon}(M_h)]
 }{
 {\cal K}_{1500,{\rm Salp}}[T,Z_{\rm Salp}(M_h)]
 }.
\end{equation}

\subsubsection{SFR/feedback factor}
At fixed $M_h$ and $z$, the SFR differs only through
$\epsilon_\star$.  Therefore
\begin{equation}
 {\cal B}_{\rm SFR}(M_h)
 =
 \frac{
 \epsilon_{\star,{\rm Chon}}[M_h,Z_{\rm Chon}(M_h)]
 }{
 \epsilon_{\star,{\rm Salp}}[M_h,Z_{\rm Salp}(M_h)]
 }.
\end{equation}
Equivalently,
\begin{equation}
 {\cal B}_{\rm SFR}(M_h)
 =
 \frac{
 v_c^2+f_wv_{s,{\rm Salp}}^2
 }{
 v_c^2+f_wv_{s,{\rm Chon}}^2[Z_{\rm Chon}(M_h)]
 }.
\end{equation}

\subsubsection{Dust-transmission factor}
We assume that dust is produced by CCSN. This is a robust assumption for $z>10$ where the evolutionary timescales for other stellar dust sources are longer than the Hubble time \citep{Todini01, dayal2022, Ferrara22a}. We take a surviving dust yield $y_d=0.1\,\Msun$ per CCSN \citep[][]{Lesniewska19, Ferrara25a}, so that
$\xi_{d,I}=y_d\nu_{{\rm SN},I}$.

Following \citet[][]{Ferrara26}, the 1500~\AA\ dust optical depth is
\begin{equation}
 \tau_{1500,I}=
 \frac{3.55}{\pi qG}
 \frac{\kappa_{1500}\xi_{d,I}}{\lambda^2}
 f_b\epsilon_0H(z)
 \frac{v_c^3}{v_c^2+f_wv_{s,I}^2},
\end{equation}
with $\kappa_{1500}=1.26\times10^5\,{\rm cm^2\,g^{-1}}$, $q=2$,
$\lambda=0.035$, and $f_b=0.158$.  Then, for mixed-slab geometry,
\begin{equation}
 T_d(\tau)=\frac{1-e^{-\tau}}{\tau}.
\end{equation}
The dust boost is
\begin{equation}
 {\cal B}_{\rm dust}(M_h)
 =
 \frac{
 T_d[\tau_{1500,{\rm Chon}}(M_h,Z_{\rm Chon})]
 }{
 T_d[\tau_{1500,{\rm Salp}}(M_h,Z_{\rm Salp})]
 }.
\end{equation}
Because the dust yield per CCSN is assumed to be IMF-independent,
\begin{equation}
 \frac{\tau_{1500,{\rm Chon}}}{\tau_{1500,{\rm Salp}}}
 =
 \frac{\nu_{{\rm SN},{\rm Chon}}}{\nu_{{\rm SN},{\rm Salp}}}
 {\cal B}_{\rm SFR}.
\end{equation}
We note that the above calculation contains some simplifying assumptions. First, the dust yield is independent of progenitor mass and metallicity; second, dust growth, { ejection, astration} and destruction are neglected. The CCSN yield is rather uncertain, and therefore more refined assumptions are not warranted. The timescale of the competing growth and destruction processes are comparable or longer than 30 Myr and hence their impact on our results is likely small. These effects are nevertheless fully included in the {\tt DELPHI} runs discussed below. 

\begin{center}
    \begin{figure*}
        \centering
    	\includegraphics[scale=0.7]{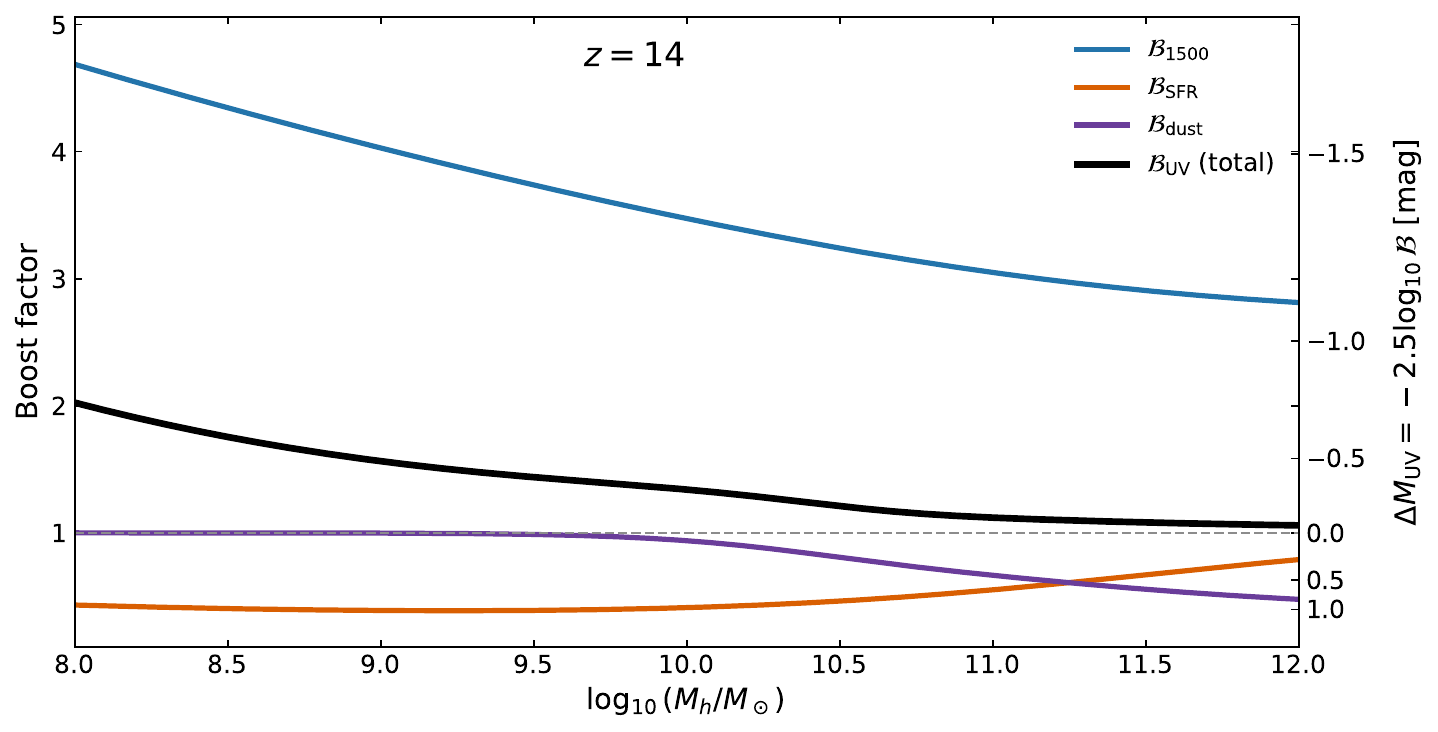}
        \caption{IMF-induced boost of the emergent 1500~\AA\ luminosity as a function of halo mass at $z=14$.  The Chon top-heavy IMF case is compared to a fixed Salpeter reference IMF, with both IMFs evaluated over $0.1<m/\Msun<300$.  The blue curve shows the direct stellar-population term, ${\cal B}_{1500}$, computed from the 30~Myr constant-SFR SB99 calibration of ${\cal K}_{1500}$.  The orange curve shows the SFR/feedback factor, ${\cal B}_{\rm SFR}$, which is below unity because the Chon IMF produces more CCSNe per unit stellar mass and therefore stronger feedback.  The purple curve shows the dust-transmission factor, ${\cal B}_{\rm dust}$, computed for the AFM mixed-slab geometry with $y_d=0.1\,\Msun$ per CCSN.  The black curve gives the total boost, ${\cal B}_{\rm UV}={\cal B}_{1500}{\cal B}_{\rm SFR}{\cal B}_{\rm dust}$. The right axis gives the corresponding magnitude offset, $\Delta M_{\rm UV}=-2.5\log_{10}{\cal B}$.  The top-heavy Chon IMF enhances the intrinsic UV output, but this is almost completely compensated by stronger feedback and, in massive halos, by enhanced dust attenuation.}
    \label{fig:fig02}
    \end{figure*}
\end{center}

\subsection{Results}
Figure~\ref{fig:fig02} shows the three factors entering the IMF-induced UV
boost, together with their product, for our fiducial calculation at $z=14$.
The direct stellar-population term, ${\cal B}_{1500}$, is the largest
individual contribution.  It decreases from ${\cal B}_{1500}\simeq 4.7$ at
$M_h=10^8\,\Msun$ to ${\cal B}_{1500}\simeq 2.8$ at
$M_h=10^{12}\,\Msun$.  This is the effect most commonly invoked \citep[e.g.][]{Trinca24, Hutter_2024, Jeong25} when arguing
that a top-heavy IMF can help explain the apparent overabundance of bright
super-early galaxies: at fixed SFR, a more top-heavy IMF produces more massive
stars and therefore a larger 1500~\AA\ luminosity.

However, this argument is incomplete if the IMF is changed self-consistently.
The same massive stars that increase ${\cal K}_{1500}$ also increase the
number of CCSNe per unit stellar mass.  This has two
consequences.  First, the larger CCSN rate raises the specific feedback energy
and lowers the star-formation efficiency at fixed halo mass.  This produces
${\cal B}_{\rm SFR}<1$ over the whole mass range, with the strongest
suppression in the low-mass, feedback-regulated regime.  Second, if each CCSN
produces a fixed surviving dust mass, the larger CCSN rate also increases the
dust column.  This has little effect in low-mass halos, where the optical depth
is small, but it becomes important at the massive end, where
${\cal B}_{\rm dust}$ drops below unity.

As a result, the net IMF-induced enhancement is much smaller than the direct
${\cal B}_{1500}$ term alone would suggest.  The total boost decreases from
${\cal B}_{\rm UV}\simeq2.0$ at $M_h=10^8\,\Msun$ to
${\cal B}_{\rm UV}\simeq1.06$ at $M_h=10^{12}\,\Msun$.  In the halo-mass range
most relevant for the bright end of the UV luminosity function,
$M_h\gtrsim10^{10}\,\Msun$, the gain is only
${\cal B}_{\rm UV}\simeq1.2$--$1.4$, corresponding to
$\Delta M_{\rm UV} \simeq -0.3$ mag.  Thus the IMF effect is certainly
not negligible, but it is too small to solve the bright-galaxy abundance
problem by itself once feedback and dust are treated consistently.

We pause to clarify the relation to the AFM interpretation.  In the AFM the main mechanism that allows bright blue galaxies to emerge is not a large increase in the intrinsic UV production efficiency, but the removal of dust attenuation by radiation-pressure-driven outflows.  In the present notation this corresponds to driving the optical depth toward $\tau_d\simeq0$, so that $T_d\to1$.  This is qualitatively different from relying on an IMF-induced boost of ${\cal K}_{1500}$: a top-heavy IMF helps, but the self-consistent
feedback and dust penalties strongly reduce its leverage.

Although Fig.~\ref{fig:fig02} is shown at $z=14$, the result is nearly
redshift-independent over the redshift range relevant for this study.  This is
not accidental.  Most of the explicit cosmological factors enter the Chon and
Salpeter calculations in the same way and therefore cancel in the boost ratios.
For example,
\begin{equation}
 {\cal B}_{\rm SFR} =
 \frac{v_c^2+f_wv_{s,{\rm Salp}}^2}
 {v_c^2+f_wv_{s,{\rm Chon}}^2},
\end{equation}
so the redshift dependence enters only through $v_c^2(M_h,z)$.  At high
redshift, $H(z)\propto(1+z)^{3/2}$ and therefore
$v_c^2\propto H^{2/3}\propto(1+z)$.  In the low-mass feedback-regulated
regime, where $v_c^2\ll f_wv_s^2$, this dependence drops out almost
completely.  In the high-mass regime, where $v_c^2\gg f_wv_s^2$, both IMFs
approach the same limiting efficiency and the ratio again tends to unity.

A similar cancellation occurs in the dust factor.  The optical depth contains
the common factor $H(z)v_c^3$ for both IMFs, while the IMF dependence enters
mainly through $\nu_{\rm SN}$ and $v_s$.  Therefore the ratio
${\cal B}_{\rm dust}$ is much less sensitive to redshift than the absolute
optical depths.  Finally, the metallicity fixed point also changes weakly with
redshift because the explicit redshift dependence in the AFM metallicity
relation is partly compensated by the redshift dependence of
$\epsilon_\star$.  The curves in Fig.~\ref{fig:fig02} should therefore be
interpreted as representative of the high-redshift regime rather than as a
special result tied to $z=14$.

We have conservatively neglected the possible contribution of PISNe to feedback and dust production.  If present, PISNe would inject substantially more energy, metals, and dust per explosion than ordinary core-collapse SNe, thereby reducing both
${\cal B}_{\rm SFR}$ and ${\cal B}_{\rm dust}$.  Their inclusion would not
directly alter the stellar-population factor ${\cal B}_{1500}$, but would
further decrease the total boost ${\cal B}_{\rm UV}$. Neglecting PISNe therefore makes our estimated UV enhancement conservative.

\begin{center}
    \begin{figure}
        \centering
    	\includegraphics[scale=0.42]{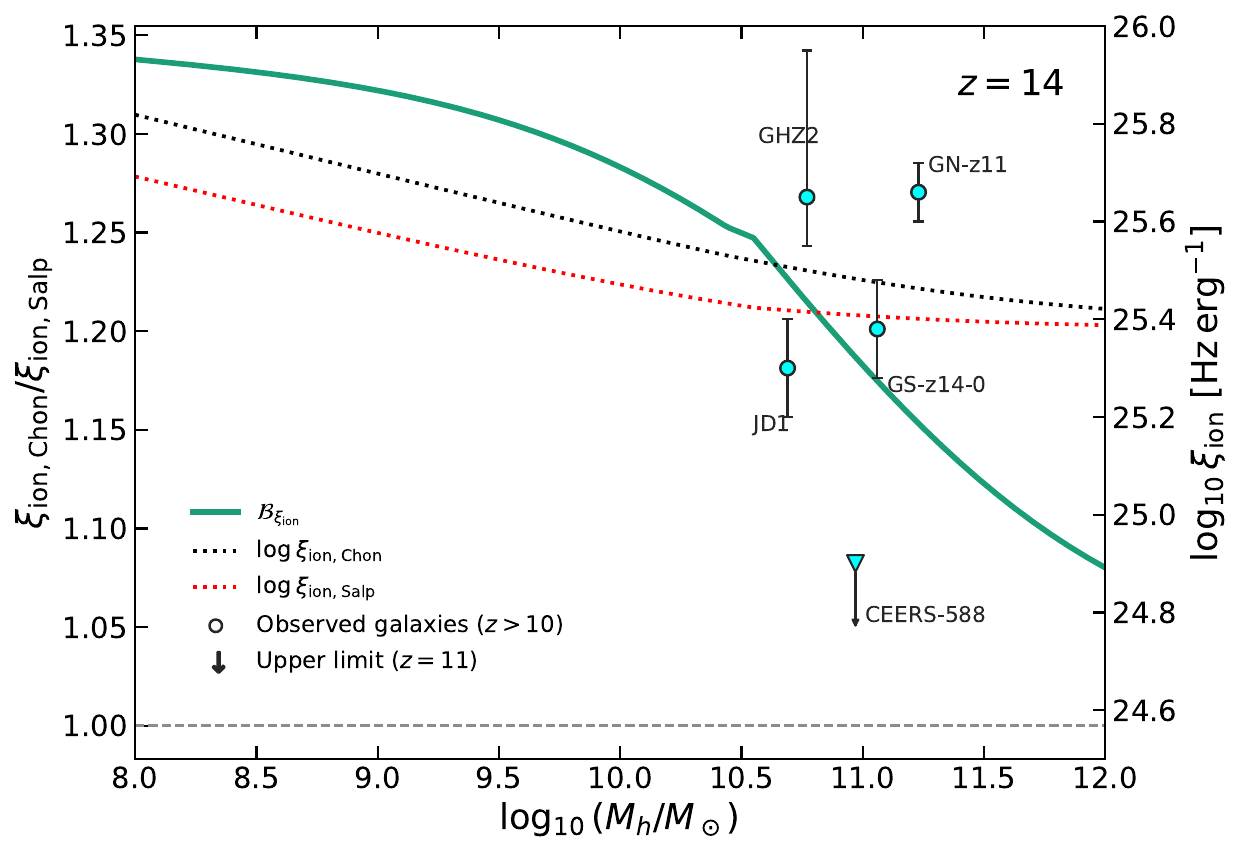}
        \caption{Ionizing-photon production efficiency for the metallicity-dependent Chon IMF relative to the fixed Salpeter reference IMF.  The green curve shows ${\cal B}_{\xi_{\rm ion}}$ as a function of halo mass at $z=14$.  The dotted curves, read on the right axis, show the corresponding absolute values of $\log_{10}\xi_{\rm ion}$ for the Chon and Salpeter cases. The Chon IMF enhances $\xi_{\rm ion}$ by only a modest factor, decreasing from $\simeq 1.3$ at low halo masses to nearly unity in the most massive halos. Data points are from \citet{Helton25,Calabro24,Hsiao24,Alvarez25,Harikane26} for GS-z14-0, GHZ2, JD1, GN-z11, CEERS-588, respectively. The halo masses are taken from \citet{Ferrara26} with metallicity computed from eq. \ref{eq:metal}. }
    \label{fig:fig03}
    \end{figure}
\end{center}

\subsubsection{Ionizing-photon production efficiency}
We also compute the IMF-dependent boost in the ionizing-photon production
efficiency,
\begin{equation}
 \xi_{\rm ion}\equiv
 \frac{\dot N_{\rm ion}}{L_\nu(1500)} ,
\end{equation}
where $\dot N_{\rm ion}$ is the production rate of H-ionizing photons and
$L_\nu(1500)$ is the intrinsic UV specific luminosity.  For consistency with
the fiducial UV calculation, both quantities are evaluated for a constant-SFR
population of age $T=30\,{\rm Myr}$, obtained by convolving the instantaneous
SB99 spectra.

\begin{center}
    \begin{figure*}
        \centering
    	\includegraphics[scale=0.12]{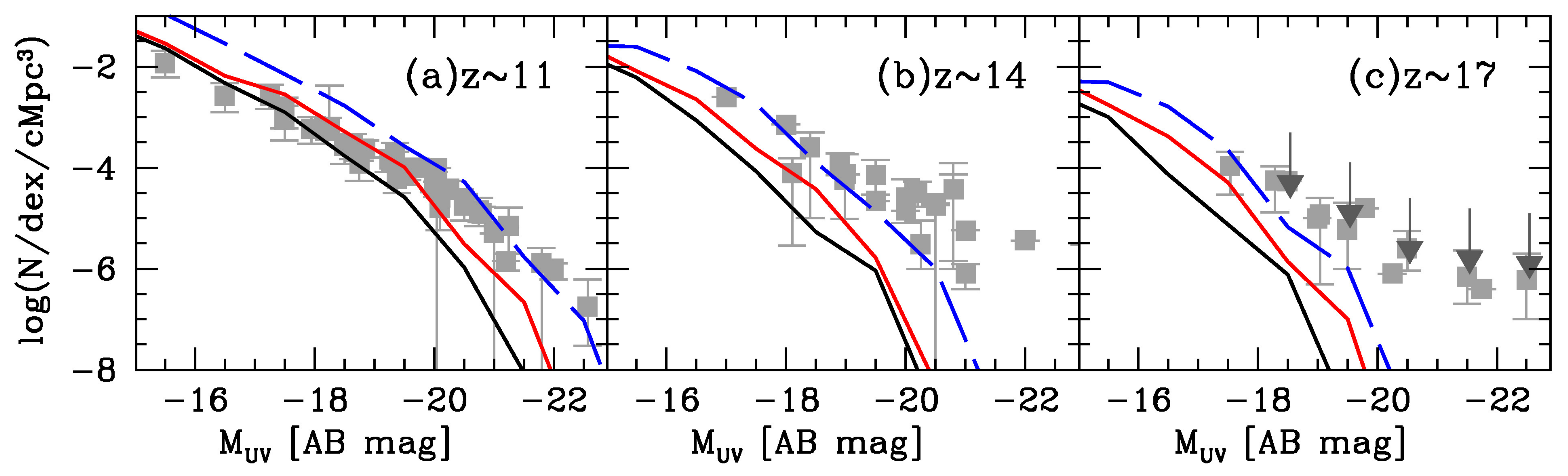}
        \caption{The redshift evolution of the UV LF at $z \sim 11-17$, as marked. In each panel, the solid black and red lines show results obtained by implementing a Salpeter and Chon IMF, respectively, in the {\sc delphi} semi-analytic model. The blue dashed line shows results assuming a Salpeter IMF for the SFR and feedback, but using the mass-to-light ratios from a Chon IMF. Points show observed data from: (a) $z \sim 11$ \citep{Donnan23a,Leung23, Perez23, Casey24, Mcleod24, asada2026, chemerynska2026}, (b) $z \sim 14$ \citep{Donnan23a, Casey24, Harikane24, Robertson24, whitler2025, asada2026,kreilgaard2026}, and (c) $z \sim 17$ \citep{Casey24, Perez25,weibel2026}. }
    \label{fig:fig04}
    \end{figure*}
\end{center}

\begin{center}
    \begin{figure}
        \centering
    	\includegraphics[scale=0.06]{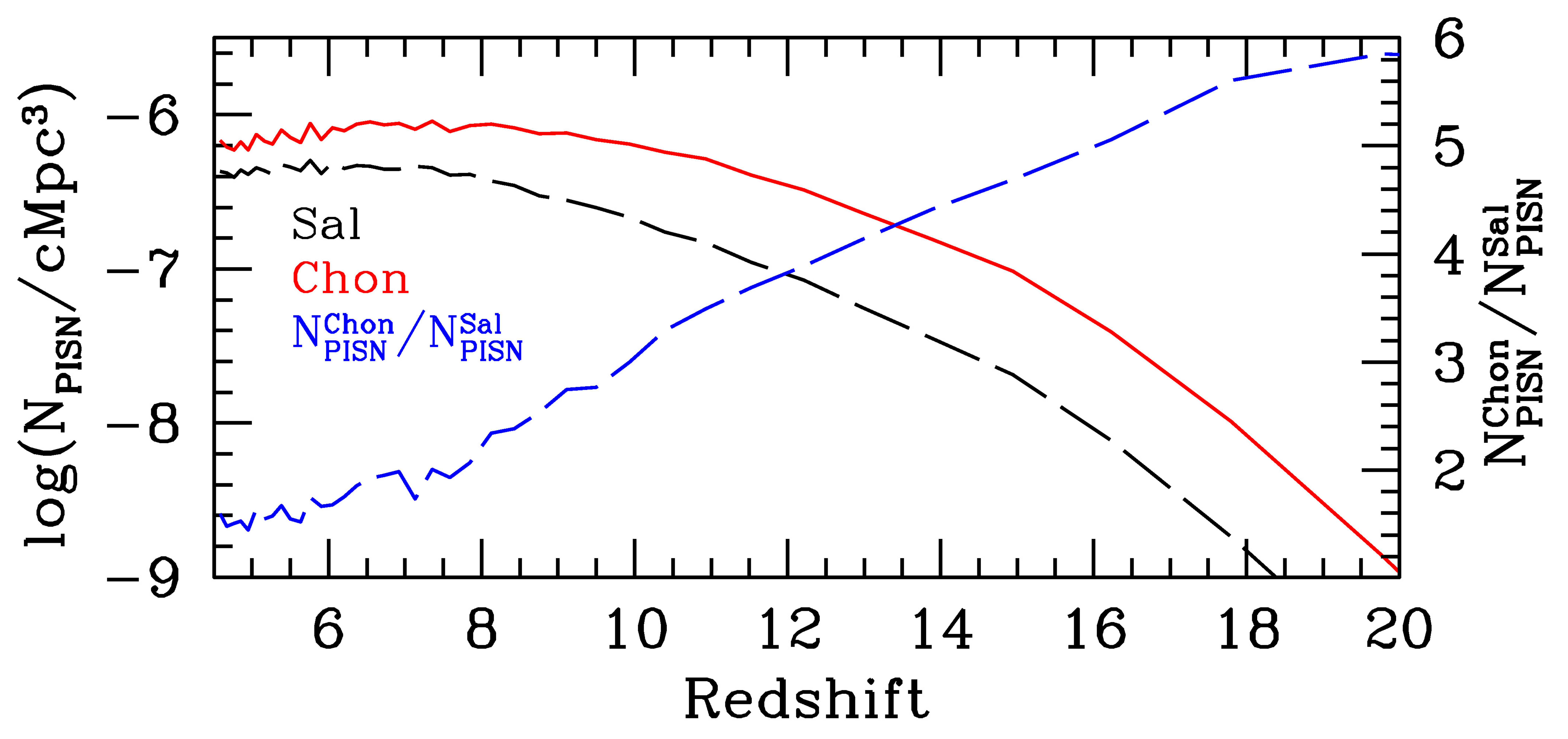}
        \caption{Redshift evolution of the PISN number density obtained from the {\sc delphi} SAM. The black dashed and red solid lines show results for the Salpeter and Chon IMFs, respectively. The blue line shows the ratio between PISN densities using the Chon versus Salpeter IMF, as marked on the right axis.}
    \label{fig:fig05}
    \end{figure}
\end{center}
Proceeding as in the previous calculations, the ionizing-efficiency boost is then defined at fixed halo mass as
\begin{equation}
 {\cal B}_{\xi_{\rm ion}}(M_h)
 =
 \frac{
 \xi_{{\rm ion},{\rm Chon}}[Z_{\rm Chon}(M_h)]
 }{
 \xi_{{\rm ion},{\rm Salp}}[Z_{\rm Salp}(M_h)]
 } .
\end{equation}
This factor isolates the change in the number of ionizing photons produced per
unit UV luminosity.  In the fiducial calculation, the Chon IMF yields
${\cal B}_{\xi_{\rm ion}}\simeq1.3$ at low halo masses, decreasing toward
unity at the highest masses as the self-consistent metallicities increase and
the two IMF-dependent spectra become more similar.

\section{Testing with simulations}
We then test the above hypotheses using the latest version of the {\sc delphi} semi-analytic model \citep{mauerhofer2025} that incorporates ISM physics - in terms of the cold gas fraction and star formation efficiency - from the {\sc sphinx} hydrodynamic simulations \citep{rosdahl2022}. 

We build the dark matter merger trees for 600 halos logarithmically equally spaced halo masses in the range $M_h = 10^{8-14} \, \Msun$ using a binary merger tree algorithm \citep{parkinson2008}. We start at $z =4.5$ and build trees up to $z \sim 40$, equally spaced by 30 Myr with a mass resolution of $M_{\rm{res}} =10^{8.0} \, \Msun$. Further, each halo at $z =4.5$ is assigned a number density by matching to the Sheth-Tormen \citep{sheth-tormen1999} halo mass function (HMF). All the progenitors of a halo are assigned the same number density as their final $z=4.5$ descendant; the resulting HMFs are found to be in accord with the Sheth-Tormen HMFs at all higher redshifts.

The starting halos of any merger tree are assigned an initial gas mass assuming a cosmological baryon-to-dark matter ratio such that $M_{\rm{g}}^{\rm i} = (\Omega_b/\Omega_m) M_h$. For halos with progenitors, the initial gas mass is the sum of the gas mass brought in by merging progenitors, and that smoothly-accreted from the intergalactic medium (IGM). For the latter, we assume smoothly accreted dark matter to drag in a cosmological ratio of baryons. 

We incorporate the cold gas fraction, $\fcold$, and star formation efficiency, $f_*^{\rm{sphinx}}$, distributions sampled from the {\sc sphinx}$^{20}$ simulation that simulates a cubic volume of 20 cMpc$^3$ using the \textsc{Ramses-RT} adaptive mesh refinement code. With a dark matter particle mass resolution  of $2.5 \times 10^5 \Msun$, this simulation resolves halos at the atomic cooling mass of $3 \times 10^7\Msun$ with 120 dark matter particles. We split the total gas mass into warm and cold ($T<1000K$) components. The cold gas mass acts as the reservoir for star formation and dust grain growth of in the ISM. The newly formed stellar mass at any time-step is calculated as
$
\Delta M_*(z) = \feff \times M_{\rm{g}}^{\rm{i}}, \label{eq:stars_formed}
$ where the ``effective" star formation efficiency is $\feff = \min(\fcold f_*^{\rm{sphinx}}, \fej)$. Here, $\fej$ is the star formation efficiency which produces enough SN energy to eject the rest of the gas \citep{dayal2014}:
\begin{equation}
\fej = \frac{v_c^2}{v_c^2 + f_w \, \nu_{\rm SN} \, E_{\rm SN}}, \label{eq:fej}
\end{equation}
where $v_c$ is the halo rotational velocity, $\nu$ is the core-collapse supernova rate, $E_{\rm SN}$ is the energy per SNII and $f_w$ is a free parameter that determines the fraction of SN energy coupling to the gas content. We consider the feedback from both SN and PISN in systems with metallicity values $Z \leq 0.1 Z_\odot$ while more enriched systems are only considered to have SN. While SNII are assumed to have an explosion energy of $E_{\rm SN}=10^{51}{\rm erg}$, PISN explode with $E_{\rm PISN} = 2 \times 10^{52}{\rm eg}$ \citep{takahashi2016}. If $\fej < f_{\rm{cold}}  f_*^{\rm{sphinx}}$, a galaxy is ``feedback-limited" i.e. it loses all its gas due to SN feedback. We fix $f_w = 4\%$, in order to match the observed redshift evolution of the UV LF at $z\leq 10$ \citep[see][]{mauerhofer2025}.

We jointly track the metal and dust contents of early galaxies including the processes of metal/dust production, astration into further star formation, ejection in outflows, dilution in (metal and dust poor) inflows of gas from the IGM, and dust destruction and growth in the ISM \citep[for details see][]{dayal2022}. We assume metals, dust and gas to be perfectly mixed in the ISM. We consider SN as the key dust factories, with each SN producing $0.1 \Msun$ of dust. We carry out the above calculations using both a Salpeter and metallicity-dependent Chon IMF between $0.1-300\msun$. For the sake of consistency with the analytic model presented above, we use the same values for metal and dust enrichment as noted in Table \ref{tab:imf_quantities}. 

Fig.~\ref{fig:fig04} shows the resulting UV LFs at $z \sim 11-17$, for both the Salpeter and Chon IMFs. For a given set of parameters, the Salpeter IMF yields the lower limit to the UV LF given it has the lowest value of the light produced per unit star formation rate. We then carry out a {\sc delphi} run using a Salpeter IMF for all other quantities, but use the UV light produced per unit SFR from the Chon IMF. This results in the blue line in the UV LF which brightens the UV magnitude by about 1.5 magnitudes, bringing the theoretical model into accord with the observed over-abundance of bright systems at these super early epochs. However, reiterating the point made above, {\it self-consistently} using SN rates, metal and dust yields, and crucially mass-to-light ratios, from the Chon IMF results in a UV LF that is different from the {\it fiducial} Salpeter one by $< 0.5$ dex, irrespective of the exact redshift.    
\section{PISN rate}
Finally, we show the redshift evolution of the number of PISN expected per unit volume from the {\sc delphi} model in Fig. \ref{fig:fig05}. As seen, the PISN density increases with decreasing redshift from $z\sim 20-7$. This is mostly driven by the increase in the star formation rate density as an increasing number of halos build up that are able to efficiently form stars. The number density saturates, and shows a slight downturn below $z \sim 7$, driven by the increasing metal enrichment of systems. Most systems at $z \sim 7$ are enriched above the $10\%\zsun$ level and therefore do not host PISN. In the context of JWST, at $z \sim 10 ~ (15)$, our Salpeter IMF model predicts $0.25 ~ (0.02)$ PISN in a typical observed volume of the order of $10^6\ {\rm cMpc^3}$. With its higher SN/PISN rate, the Chon IMF results in 3-5 times more PISN, with values of $0.7 ~ (0.1)$ at $z \sim 10 ~ (15)$. We also show the ratio of PISN density for the Chon versus the Salpeter IMF in the same figure. While the Chon IMF predicts twice as many PISN per unit volume as Salpeter at $z \sim 8$, the value rises to a factor 6 by $z \sim 20$. We defer a full calculation on the visibility of such PISN to a future work. We conclude that while different IMFs have very limited impact on the UV LF, such degeneracy can be broken by the detection of PISN whose rate significantly differs between the two IMFs considered here.  

\section{Summary}

We have investigated whether a metallicity-dependent, top-heavy IMF can
explain the unexpectedly large abundance of UV-bright galaxies at cosmic
dawn.  We compared the physically motivated top-heavy Chon IMF to a fixed Salpeter IMF over the common mass range $0.1<m/\Msun<300$, consistently accounting for the IMF dependence of the UV luminosity, SN feedback, metal enrichment, and dust production.  We
first developed an analytic calculation in which the gas metallicity and IMF
are solved self-consistently at fixed halo mass, and then tested the results
with the {\sc delphi} semi-analytic model.  Our main findings are:

\begin{itemize}

\item[\color{red}$\blacksquare$] At fixed SFR, the Chon IMF substantially
increases the intrinsic UV luminosity: ${\cal B}_{1500}$ decreases from
$\simeq4.7$ at $M_h=10^8\,\Msun$ to $\simeq2.8$ at
$M_h=10^{12}\,\Msun$.  The enhancement of the ionizing-photon production
efficiency is much smaller, with ${\cal B}_{\xi_{\rm ion}}\simeq1.3$ in
low-mass halos and approaching unity at the massive end.

\item[\color{red}$\blacksquare$] The same massive stars that enhance the UV
output also increase the SN rate, thereby strengthening feedback and dust
attenuation.  Once these effects are included, the net boost falls from
${\cal B}_{\rm UV}\simeq2.0$ at $M_h=10^8\,\Msun$ to $\simeq1.06$ at
$10^{12}\,\Msun$.  For the halos most relevant to the bright end of the UV
luminosity function, $M_h\gtrsim10^{10}\,\Msun$, the enhancement is only
${\cal B}_{\rm UV}\simeq1.2$--$1.4$, or $\Delta M_{\rm UV}\simeq-0.3$ mag.
These boost factors depend only weakly on redshift throughout the regime of
interest.

\item[\color{red}$\blacksquare$] The {\sc delphi} calculations confirm this
compensation.  Replacing only the Salpeter mass-to-light ratio by the Chon one
brightens galaxies by about $1.5$ mag and can reproduce the observed excess of
bright sources.  However, when the Chon CCSN rates, feedback, enrichment,
dust production, and mass-to-light ratios are all used self-consistently, the
predicted UV luminosity function differs from the Salpeter result by less than
$0.5$ dex over $z\simeq11$--$17$.

\item[\color{red}$\blacksquare$] PISNe provide a more sensitive diagnostic of
the IMF.  In a survey volume of $10^6\,{\rm cMpc^3}$, the Salpeter model
predicts $0.25\ (0.02)$ PISNe at $z\simeq10\ (15)$, whereas the Chon model
predicts $0.7\ (0.1)$.  The Chon-to-Salpeter PISN abundance ratio increases
from about $2$ at $z\simeq8$ to about $6$ at $z\simeq20$.

\end{itemize}

We conclude that the direct luminosity gain of a top-heavy IMF cannot be
considered independently of its dynamical and chemical consequences.  Once
feedback and dust are treated consistently, a top-heavy IMF alone is unable
to explain the abundance of UV-bright galaxies at cosmic dawn.  
Conversely, because the predicted PISN abundance is
far more sensitive to the high-mass IMF, future PISN searches may offer a
cleaner route to distinguishing between these different IMFs.

\section*{Acknowledgments}
AF gratefully acknowledges the generous support and hospitality of the Canadian Institute for Theoretical Astrophysics at the University of Toronto, where part of this work was carried out under a Visiting Professor Fellowship. P. Dayal warmly acknowledges support from an NSERC discovery grant (RGPIN-2025-06182).

\bibliographystyle{mn2e}


\bibliography{oja_template}

@article{Jeong25,
doi = {10.3847/1538-4357/ada27d},
url = {https://doi.org/10.3847/1538-4357/ada27d},
year = {2025},
month = {jan},
publisher = {The American Astronomical Society},
volume = {980},
number = {1},
pages = {10},
author = {Jeong, Tae Bong and Jeon, Myoungwon and Song, Hyunmi and Bromm, Volker},
title = {Simulating High-redshift Galaxies: Enhancing UV Luminosity with Star Formation Efficiency and a Top-heavy IMF},
journal = {The Astrophysical Journal}
}

@ARTICLE{Alvarez25,
       author = {{{\'A}lvarez-M{\'a}rquez}, J. and {Crespo G{\'o}mez}, A. and {Colina}, L. and {Langeroodi}, D. and {Marques-Chaves}, R. and {Prieto-Jim{\'e}nez}, C. and {Bik}, A. and {Alonso-Herrero}, A. and {Boogaard}, L. and {Costantin}, L. and {Garc{\'\i}a-Mar{\'\i}n}, M. and {Gillman}, S. and {Hjorth}, J. and {Iani}, E. and {Jermann}, I. and {Labiano}, A. and {Melinder}, J. and {Meyer}, R. and {{\"O}stlin}, G. and {P{\'e}rez-Gonz{\'a}lez}, P.~G. and {Rinaldi}, P. and {Walter}, F. and {van der Werf}, P. and {Wright}, G.},
        title = "{Insight into the starburst nature of Galaxy GN-z11 with JWST MIRI spectroscopy}",
      journal = {\aap},
         year = 2025,
        month = mar,
       volume = {695},
          eid = {A250},
        pages = {A250},
          doi = {10.1051/0004-6361/202451731},
archivePrefix = {arXiv},
       eprint = {2412.12826},
 primaryClass = {astro-ph.GA},
       adsurl = {https://ui.adsabs.harvard.edu/abs/2025A&A...695A.250A}
}

@ARTICLE{asada2026,
       author = {{Asada}, Yoshihisa and {Willott}, Chris J. and {Muzzin}, Adam and {Brada{\v{c}}}, Maru{\v{s}}a and {Brammer}, Gabriel and {Desprez}, Guillaume and {Iyer}, Kartheik G. and {Marchesini}, Danilo and {Martis}, Nicholas S. and {Noirot}, Ga{\"e}l and {Sarrouh}, Ghassan T.~E. and {Sawicki}, Marcin and {Withers}, Sunna and {Fujimoto}, Seiji and {Felicioni}, Giordano and {Goovaerts}, Ilias and {Jude{\v{z}}}, Jon and {Jagga}, Naadiyah and {Merchant}, Maya and {M{\'e}rida}, Rosa M. and {Robbins}, Luke},
        title = "{Earliest Galaxy Evolution in the CANUCS+Technicolor Fields: Galaxy Properties at z {\ensuremath{\sim}} 10─16 Seen with the Full NIRCam Medium- and Broadband Filters}",
      journal = {\apj},
         year = 2026,
        month = jan,
       volume = {996},
       number = {2},
          eid = {115},
        pages = {115},
          doi = {10.3847/1538-4357/ae1f8d},
archivePrefix = {arXiv},
       eprint = {2507.03124},
 primaryClass = {astro-ph.GA},
       adsurl = {https://ui.adsabs.harvard.edu/abs/2026ApJ...996..115A}
}

@ARTICLE{Bromm01,
       author = {{Bromm}, V. and {Ferrara}, A. and {Coppi}, P.~S. and {Larson}, R.~B.},
        title = "{The fragmentation of pre-enriched primordial objects}",
      journal = {\mnras},
         year = 2001,
        month = dec,
       volume = {328},
       number = {3},
        pages = {969-976},
          doi = {10.1046/j.1365-8711.2001.04915.x},
archivePrefix = {arXiv},
       eprint = {astro-ph/0104271},
 primaryClass = {astro-ph},
       adsurl = {https://ui.adsabs.harvard.edu/abs/2001MNRAS.328..969B}
}

@ARTICLE{Bunker23,
        author = {{Bunker}, Andrew J. and {Saxena}, Aayush and {Cameron}, Alex J. and {Willott}, Chris J. and {Curtis-Lake}, Emma and {Jakobsen}, Peter and {Carniani}, Stefano and {Smit}, Renske and {Maiolino}, Roberto and {Witstok}, Joris and {Curti}, Mirko and {D'Eugenio}, Francesco and {Jones}, Gareth C. and {Ferruit}, Pierre and {Arribas}, Santiago and {Charlot}, Stephane and {Chevallard}, Jacopo and {Giardino}, Giovanna and {de Graaff}, Anna and {Looser}, Tobias J. and {L{\"u}tzgendorf}, Nora and {Maseda}, Michael V. and {Rawle}, Tim and {Rix}, Hans-Walter and {Del Pino}, Bruno Rodr{\'\i}guez and {Alberts}, Stacey and {Egami}, Eiichi and {Eisenstein}, Daniel J. and {Endsley}, Ryan and {Hainline}, Kevin and {Hausen}, Ryan and {Johnson}, Benjamin D. and {Rieke}, George and {Rieke}, Marcia and {Robertson}, Brant E. and {Shivaei}, Irene and {Stark}, Daniel P. and {Sun}, Fengwu and {Tacchella}, Sandro and {Tang}, Mengtao and {Williams}, Christina C. and {Willmer}, Christopher N.~A. and {Baker}, William M. and {Baum}, Stefi and {Bhatawdekar}, Rachana and {Bowler}, Rebecca and {Boyett}, Kristan and {Chen}, Zuyi and {Circosta}, Chiara and {Helton}, Jakob M. and {Ji}, Zhiyuan and {Kumari}, Nimisha and {Lyu}, Jianwei and {Nelson}, Erica and {Parlanti}, Eleonora and {Perna}, Michele and {Sandles}, Lester and {Scholtz}, Jan and {Suess}, Katherine A. and {Topping}, Michael W. and {{\"U}bler}, Hannah and {Wallace}, Imaan E.~B. and {Whitler}, Lily},
        title = "{JADES NIRSpec Spectroscopy of GN-z11: Lyman-{\ensuremath{\alpha}} emission and possible enhanced nitrogen abundance in a z = 10.60 luminous galaxy}",
      journal = {\aap},
         year = 2023,
        month = sep,
       volume = {677},
          eid = {A88},
        pages = {A88},
          doi = {10.1051/0004-6361/202346159},
archivePrefix = {arXiv},
       eprint = {2302.07256},
 primaryClass = {astro-ph.GA},
       adsurl = {https://ui.adsabs.harvard.edu/abs/2023A&A...677A..88B}
}

@ARTICLE{Calabro24,
       author = {{Calabr{\`o}}, Antonello and {Castellano}, Marco and {Zavala}, Jorge A. and {Pentericci}, Laura and {Arrabal Haro}, Pablo and {Bakx}, Tom J.~L.~C. and {Burgarella}, Denis and {Casey}, Caitlin M. and {Dickinson}, Mark and {Finkelstein}, Steven L. and {Fontana}, Adriano and {Llerena}, Mario and {Mascia}, Sara and {Merlin}, Emiliano and {Mitsuhashi}, Ikki and {Napolitano}, Lorenzo and {Paris}, Diego and {P{\'e}rez-Gonz{\'a}lez}, Pablo G. and {Roberts-Borsani}, Guido and {Santini}, Paola and {Treu}, Tommaso and {Vanzella}, Eros},
        title = "{Evidence of Extreme Ionization Conditions and Low Metallicity in GHZ2/GLASS-Z12 from a Combined Analysis of NIRSpec and MIRI Observations}",
      journal = {\apj},
         year = 2024,
        month = nov,
       volume = {975},
       number = {2},
          eid = {245},
        pages = {245},
          doi = {10.3847/1538-4357/ad7602},
archivePrefix = {arXiv},
       eprint = {2403.12683},
 primaryClass = {astro-ph.GA},
       adsurl = {https://ui.adsabs.harvard.edu/abs/2024ApJ...975..245C}
}

@ARTICLE{Cameron24,
       author = {{Cameron}, Alex J. and {Katz}, Harley and {Witten}, Callum and {Saxena}, Aayush and {Laporte}, Nicolas and {Bunker}, Andrew J.},
        title = "{Nebular dominated galaxies: insights into the stellar initial mass function at high redshift}",
      journal = {\mnras},
         year = 2024,
        month = oct,
       volume = {534},
       number = {1},
        pages = {523-543},
          doi = {10.1093/mnras/stae1547},
archivePrefix = {arXiv},
       eprint = {2311.02051},
 primaryClass = {astro-ph.GA},
       adsurl = {https://ui.adsabs.harvard.edu/abs/2024MNRAS.534..523C}
}

@ARTICLE{Carniani24a,
       author = {{Carniani}, Stefano and {Hainline}, Kevin and {D'Eugenio}, Francesco and {Eisenstein}, Daniel J. and {Jakobsen}, Peter and {Witstok}, Joris and {Johnson}, Benjamin D. and {Chevallard}, Jacopo and {Maiolino}, Roberto and {Helton}, Jakob M. and {Willott}, Chris and {Robertson}, Brant and {Alberts}, Stacey and {Arribas}, Santiago and {Baker}, William M. and {Bhatawdekar}, Rachana and {Boyett}, Kristan and {Bunker}, Andrew J. and {Cameron}, Alex J. and {Cargile}, Phillip A. and {Charlot}, St{\'e}phane and {Curti}, Mirko and {Curtis-Lake}, Emma and {Egami}, Eiichi and {Giardino}, Giovanna and {Isaak}, Kate and {Ji}, Zhiyuan and {Jones}, Gareth C. and {Kumari}, Nimisha and {Maseda}, Michael V. and {Parlanti}, Eleonora and {P{\'e}rez-Gonz{\'a}lez}, Pablo G. and {Rawle}, Tim and {Rieke}, George and {Rieke}, Marcia and {Del Pino}, Bruno Rodr{\'\i}guez and {Saxena}, Aayush and {Scholtz}, Jan and {Smit}, Renske and {Sun}, Fengwu and {Tacchella}, Sandro and {{\"U}bler}, Hannah and {Venturi}, Giacomo and {Williams}, Christina C. and {Willmer}, Christopher N.~A.},
        title = "{Spectroscopic confirmation of two luminous galaxies at a redshift of 14}",
      journal = {\nat},
         year = 2024,
        month = sep,
       volume = {633},
       number = {8029},
        pages = {318-322},
          doi = {10.1038/s41586-024-07860-9},
archivePrefix = {arXiv},
       eprint = {2405.18485},
 primaryClass = {astro-ph.GA},
       adsurl = {https://ui.adsabs.harvard.edu/abs/2024Natur.633..318C}
}

@ARTICLE{Casey24,
       author = {{Casey}, Caitlin M. and {Akins}, Hollis B. and {Shuntov}, Marko and {Ilbert}, Olivier and {Paquereau}, Louise and {Franco}, Maximilien and {Hayward}, Christopher C. and {Finkelstein}, Steven L. and {Boylan-Kolchin}, Michael and {Robertson}, Brant E. and {Allen}, Natalie and {Brinch}, Malte and {Cooper}, Olivia R. and {Ding}, Xuheng and {Drakos}, Nicole E. and {Faisst}, Andreas L. and {Fujimoto}, Seiji and {Gillman}, Steven and {Harish}, Santosh and {Hirschmann}, Michaela and {Jin}, Shuowen and {Kartaltepe}, Jeyhan S. and {Koekemoer}, Anton M. and {Kokorev}, Vasily and {Liu}, Daizhong and {Long}, Arianna S. and {Magdis}, Georgios and {Maraston}, Claudia and {Martin}, Crystal L. and {McCracken}, Henry Joy and {McKinney}, Jed and {Mobasher}, Bahram and {Rhodes}, Jason and {Rich}, R. Michael and {Sanders}, David B. and {Silverman}, John D. and {Toft}, Sune and {Vijayan}, Aswin P. and {Weaver}, John R. and {Wilkins}, Stephen M. and {Yang}, Lilan and {Zavala}, Jorge A.},
        title = "{COSMOS-Web: Intrinsically Luminous z {\ensuremath{\gtrsim}} 10 Galaxy Candidates Test Early Stellar Mass Assembly}",
      journal = {\apj},
         year = 2024,
        month = apr,
       volume = {965},
       number = {1},
          eid = {98},
        pages = {98},
          doi = {10.3847/1538-4357/ad2075},
archivePrefix = {arXiv},
       eprint = {2308.10932},
 primaryClass = {astro-ph.GA},
       adsurl = {https://ui.adsabs.harvard.edu/abs/2024ApJ...965...98C}
}

@ARTICLE{Castellano23,
       author = {{Castellano}, Marco and {Fontana}, Adriano and {Treu}, Tommaso and {Merlin}, Emiliano and {Santini}, Paola and {Bergamini}, Pietro and {Grillo}, Claudio and {Rosati}, Piero and {Acebron}, Ana and {Leethochawalit}, Nicha and {Paris}, Diego and {Bonchi}, Andrea and {Belfiori}, Davide and {Calabr{\`o}}, Antonello and {Correnti}, Matteo and {Nonino}, Mario and {Polenta}, Gianluca and {Trenti}, Michele and {Boyett}, Kristan and {Brammer}, G. and {Broadhurst}, Tom and {Caminha}, Gabriel B. and {Chen}, Wenlei and {Filippenko}, Alexei V. and {Fortuni}, Flaminia and {Glazebrook}, Karl and {Mascia}, Sara and {Mason}, Charlotte A. and {Menci}, Nicola and {Meneghetti}, Massimo and {Mercurio}, Amata and {Metha}, Benjamin and {Morishita}, Takahiro and {Nanayakkara}, Themiya and {Pentericci}, Laura and {Roberts-Borsani}, Guido and {Roy}, Namrata and {Vanzella}, Eros and {Vulcani}, Benedetta and {Yang}, Lilan and {Wang}, Xin},
        title = "{Early Results from GLASS-JWST. XIX. A High Density of Bright Galaxies at z {\ensuremath{\approx}} 10 in the A2744 Region}",
      journal = {\apjl},
         year = 2023,
        month = may,
       volume = {948},
       number = {2},
          eid = {L14},
        pages = {L14},
          doi = {10.3847/2041-8213/accea5},
archivePrefix = {arXiv},
       eprint = {2212.06666},
 primaryClass = {astro-ph.GA},
       adsurl = {https://ui.adsabs.harvard.edu/abs/2023ApJ...948L..14C}
}

@ARTICLE{chemerynska2026,
       author = {{Chemerynska}, Iryna and {Atek}, Hakim and {Furtak}, Lukas J. and {Chisholm}, John and {Endsley}, Ryan and {Kokorev}, Vasily and {Rosdahl}, Joki and {Blaizot}, Jeremy and {Adamo}, Angela and {Bouwens}, Rychard and {Fujimoto}, Seiji and {Korber}, Damien and {Mason}, Charlotte and {McQuinn}, Kristen B.~W. and {Mu{\~n}oz}, Julian B. and {Natarajan}, Priyamvada and {Nelson}, Erica and {Oesch}, Pascal A. and {Pan}, Richard and {Richard}, Johan and {Saldana-Lopez}, Alberto and {Schaerer}, Daniel and {Volonteri}, Marta and {Zitrin}, Adi and {Berg}, Danielle A. and {Claeyssens}, Ad{\'e}la{\"\i}de and {Dessauges-Zavadsky}, Miroslava and {Jecmen}, Michelle and {Labb{\'e}}, Ivo and {Naidu}, Rohan and {Trebitsch}, Maxime},
        title = "{The first GLIMPSE of the faint galaxy population at Cosmic Dawn with JWST: The evolution of the ultraviolet luminosity function across z {\ensuremath{\sim}} 9{\ensuremath{-}}15}",
      journal = {\mnras},
         year = 2026,
        month = feb,
       volume = {546},
       number = {2},
          doi = {10.1093/mnras/staf2267},
archivePrefix = {arXiv},
       eprint = {2509.24881},
 primaryClass = {astro-ph.GA},
       adsurl = {https://ui.adsabs.harvard.edu/abs/2026MNRAS.546f2267C}
}

@ARTICLE{Chon21,
       author = {{Chon}, Sunmyon and {Omukai}, Kazuyuki and {Schneider}, Raffaella},
        title = "{Transition of the initial mass function in the metal-poor environments}",
      journal = {\mnras},
         year = 2021,
        month = dec,
       volume = {508},
       number = {3},
        pages = {4175-4192},
          doi = {10.1093/mnras/stab2497},
archivePrefix = {arXiv},
       eprint = {2103.04997},
 primaryClass = {astro-ph.GA},
       adsurl = {https://ui.adsabs.harvard.edu/abs/2021MNRAS.508.4175C}
}

@ARTICLE{Chon22,
       author = {{Chon}, Sunmyon and {Ono}, Haruka and {Omukai}, Kazuyuki and {Schneider}, Raffaella},
        title = "{Impact of the cosmic background radiation on the initial mass function of metal-poor stars}",
      journal = {\mnras},
         year = 2022,
        month = aug,
       volume = {514},
       number = {3},
        pages = {4639-4654},
          doi = {10.1093/mnras/stac1549},
archivePrefix = {arXiv},
       eprint = {2205.15328},
 primaryClass = {astro-ph.GA},
       adsurl = {https://ui.adsabs.harvard.edu/abs/2022MNRAS.514.4639C}
}

@ARTICLE{Chon24,
       author = {{Chon}, Sunmyon and {Hosokawa}, Takashi and {Omukai}, Kazuyuki and {Schneider}, Raffaella},
        title = "{Impact of radiative feedback on the initial mass function of metal-poor stars}",
      journal = {\mnras},
         year = 2024,
        month = may,
       volume = {530},
       number = {3},
        pages = {2453-2474},
          doi = {10.1093/mnras/stae1027},
archivePrefix = {arXiv},
       eprint = {2312.13339},
 primaryClass = {astro-ph.GA},
       adsurl = {https://ui.adsabs.harvard.edu/abs/2024MNRAS.530.2453C}
}

@ARTICLE{Cueto23,
       author = {{Cueto}, Elie R. and {Hutter}, Anne and {Dayal}, Pratika and {Gottl{\"o}ber}, Stefan and {Heintz}, Kasper E. and {Mason}, Charlotte and {Trebitsch}, Maxime and {Yepes}, Gustavo},
        title = "{ASTRAEUS. IX. Impact of an evolving stellar initial mass function on early galaxies and reionisation}",
      journal = {\aap},
         year = 2024,
        month = jun,
       volume = {686},
          eid = {A138},
        pages = {A138},
          doi = {10.1051/0004-6361/202349017},
archivePrefix = {arXiv},
       eprint = {2312.12109},
 primaryClass = {astro-ph.GA},
       adsurl = {https://ui.adsabs.harvard.edu/abs/2024A&A...686A.138C}
}

@ARTICLE{Cueto26,
       author = {{Cueto}, Elie and {Hutter}, Anne},
        title = "{Bright Galaxies at Cosmic Dawn: A Cloud-Scale Star Formation Model Unifying Variable SFE, IMFs, and Stochasticity}",
      journal = {arXiv e-prints},
         year = 2026,
        month = jul,
          eid = {arXiv:2607.18868},
        pages = {arXiv:2607.18868},
          doi = {10.48550/arXiv.2607.18868},
archivePrefix = {arXiv},
       eprint = {2607.18868},
 primaryClass = {astro-ph.GA},
       adsurl = {https://ui.adsabs.harvard.edu/abs/2026arXiv260718868C}
}

@ARTICLE{Curtis23,
       author = {{Curtis-Lake}, Emma and {Carniani}, Stefano and {Cameron}, Alex and {Charlot}, Stephane and {Jakobsen}, Peter and {Maiolino}, Roberto and {Bunker}, Andrew and {Witstok}, Joris and {Smit}, Renske and {Chevallard}, Jacopo and {Willott}, Chris and {Ferruit}, Pierre and {Arribas}, Santiago and {Bonaventura}, Nina and {Curti}, Mirko and {D'Eugenio}, Francesco and {Franx}, Marijn and {Giardino}, Giovanna and {Looser}, Tobias J. and {L{\"u}tzgendorf}, Nora and {Maseda}, Michael V. and {Rawle}, Tim and {Rix}, Hans-Walter and {Rodr{\'\i}guez del Pino}, Bruno and {{\"U}bler}, Hannah and {Sirianni}, Marco and {Dressler}, Alan and {Egami}, Eiichi and {Eisenstein}, Daniel J. and {Endsley}, Ryan and {Hainline}, Kevin and {Hausen}, Ryan and {Johnson}, Benjamin D. and {Rieke}, Marcia and {Robertson}, Brant and {Shivaei}, Irene and {Stark}, Daniel P. and {Tacchella}, Sandro and {Williams}, Christina C. and {Willmer}, Christopher N.~A. and {Bhatawdekar}, Rachana and {Bowler}, Rebecca and {Boyett}, Kristan and {Chen}, Zuyi and {de Graaff}, Anna and {Helton}, Jakob M. and {Hviding}, Raphael E. and {Jones}, Gareth C. and {Kumari}, Nimisha and {Lyu}, Jianwei and {Nelson}, Erica and {Perna}, Michele and {Sandles}, Lester and {Saxena}, Aayush and {Suess}, Katherine A. and {Sun}, Fengwu and {Topping}, Michael W. and {Wallace}, Imaan E.~B. and {Whitler}, Lily},
        title = "{Spectroscopic confirmation of four metal-poor galaxies at z = 10.3-13.2}",
      journal = {Nature Astronomy},
         year = 2023,
        month = apr,
          doi = {10.1038/s41550-023-01918-w},
archivePrefix = {arXiv},
       eprint = {2212.04568},
 primaryClass = {astro-ph.GA},
       adsurl = {https://ui.adsabs.harvard.edu/abs/2023NatAs.tmp...66C}
}

@ARTICLE{Dayal18,
       author = {{Dayal}, Pratika and {Ferrara}, Andrea},
        title = "{Early galaxy formation and its large-scale effects}",
      journal = {\physrep},
         year = 2018,
        month = dec,
       volume = {780},
        pages = {1-64},
          doi = {10.1016/j.physrep.2018.10.002},
archivePrefix = {arXiv},
       eprint = {1809.09136},
 primaryClass = {astro-ph.GA},
       adsurl = {https://ui.adsabs.harvard.edu/abs/2018PhR...780....1D}
}

@ARTICLE{dayal2014,
       author = {{Dayal}, Pratika and {Ferrara}, Andrea and {Dunlop}, James S. and {Pacucci}, Fabio},
        title = "{Essential physics of early galaxy formation}",
      journal = {\mnras},
         year = 2014,
        month = dec,
       volume = {445},
       number = {3},
        pages = {2545-2557},
          doi = {10.1093/mnras/stu1848},
archivePrefix = {arXiv},
       eprint = {1405.4862},
 primaryClass = {astro-ph.GA},
       adsurl = {https://ui.adsabs.harvard.edu/abs/2014MNRAS.445.2545D}
}

@ARTICLE{dayal2022,
       author = {{Dayal}, P. and {Ferrara}, A. and {Sommovigo}, L. and {Bouwens}, R. and {Oesch}, P.~A. and {Smit}, R. and {Gonzalez}, V. and {Schouws}, S. and {Stefanon}, M. and {Kobayashi}, C. and {Bremer}, J. and {Algera}, H.~S.~B. and {Aravena}, M. and {Bowler}, R.~A.~A. and {da Cunha}, E. and {Fudamoto}, Y. and {Graziani}, L. and {Hodge}, J. and {Inami}, H. and {De Looze}, I. and {Pallottini}, A. and {Riechers}, D. and {Schneider}, R. and {Stark}, D. and {Endsley}, R.},
        title = "{The ALMA REBELS survey: the dust content of z   7 Lyman break galaxies}",
      journal = {\mnras},
         year = 2022,
        month = may,
       volume = {512},
       number = {1},
        pages = {989-1002},
          doi = {10.1093/mnras/stac537},
archivePrefix = {arXiv},
       eprint = {2202.11118},
 primaryClass = {astro-ph.GA},
       adsurl = {https://ui.adsabs.harvard.edu/abs/2022MNRAS.512..989D}
}

@ARTICLE{Donnan23a,
       author = {{Donnan}, C.~T. and {McLeod}, D.~J. and {Dunlop}, J.~S. and {McLure}, R.~J. and {Carnall}, A.~C. and {Begley}, R. and {Cullen}, F. and {Hamadouche}, M.~L. and {Bowler}, R.~A.~A. and {Magee}, D. and {McCracken}, H.~J. and {Milvang-Jensen}, B. and {Moneti}, A. and {Targett}, T.},
        title = "{The evolution of the galaxy UV luminosity function at redshifts z ≃ 8 - 15 from deep JWST and ground-based near-infrared imaging}",
      journal = {\mnras},
         year = 2023,
        month = feb,
       volume = {518},
       number = {4},
        pages = {6011-6040},
          doi = {10.1093/mnras/stac3472},
archivePrefix = {arXiv},
       eprint = {2207.12356},
 primaryClass = {astro-ph.GA},
       adsurl = {https://ui.adsabs.harvard.edu/abs/2023MNRAS.518.6011D}
}

@ARTICLE{Durrant26,
       author = {{Durrant}, Anna and {Crain}, Robert A. and {Lacey}, Cedric G. and {Schaye}, Joop and {Smit}, Renske and {Gebek}, Andrea and {Schaller}, Matthieu and {Lu}, Shengdong and {Chaikin}, Evgenii and {Andreadis}, Nick and {Baes}, Maarten and {Bate}, Matthew R. and {Ben{\'\i}tez-Llambay}, Alejandro and {Frenk}, Carlos S. and {Hu{\v{s}}ko}, Filip and {McGibbon}, Robert J. and {Ploeckinger}, Sylvia and {Richings}, Alexander J.},
        title = "{Cosmological simulations of the high-redshift galaxy population adopting a variable stellar initial mass function}",
      journal = {arXiv e-prints},
         year = 2026,
        month = jul,
          eid = {arXiv:2607.16404},
        pages = {arXiv:2607.16404},
          doi = {10.48550/arXiv.2607.16404},
archivePrefix = {arXiv},
       eprint = {2607.16404},
 primaryClass = {astro-ph.GA},
       adsurl = {https://ui.adsabs.harvard.edu/abs/2026arXiv260716404D}
}

@inproceedings{Elmegreen04,
  author    = {Elmegreen, Bruce G.},
  title     = {The IMF in Starbursts},
  booktitle = {Starbursts: From 30 Doradus to Lyman Break Galaxies},
  year      = {2004},
  eprint    = {astro-ph/0411193},
  archivePrefix = {arXiv}
}

@ARTICLE{Ferrara22a,
       author = {{Ferrara}, A. and {Sommovigo}, L. and {Dayal}, P. and {Pallottini}, A. and {Bouwens}, R.~J. and {Gonzalez}, V. and {Inami}, H. and {Smit}, R. and {Bowler}, R.~A.~A. and {Endsley}, R. and {Oesch}, P. and {Schouws}, S. and {Stark}, D. and {Stefanon}, M. and {Aravena}, M. and {da Cunha}, E. and {De Looze}, I. and {Fudamoto}, Y. and {Graziani}, L. and {Hodge}, J. and {Riechers}, D. and {Schneider}, R. and {Algera}, H.~S.~B. and {Barrufet}, L. and {Hygate}, A.~P.~S. and {Labb{\'e}}, I. and {Li}, C. and {Nanayakkara}, T. and {Topping}, M. and {van der Werf}, P.},
        title = "{The ALMA REBELS Survey. Epoch of Reionization giants: Properties of dusty galaxies at z ≍ 7}",
      journal = {\mnras},
         year = 2022,
        month = may,
       volume = {512},
       number = {1},
        pages = {58-72},
          doi = {10.1093/mnras/stac460},
archivePrefix = {arXiv},
       eprint = {2202.07666},
 primaryClass = {astro-ph.GA},
       adsurl = {https://ui.adsabs.harvard.edu/abs/2022MNRAS.512...58F}
}

@ARTICLE{Ferrara23,
       author = {{Ferrara}, Andrea and {Pallottini}, Andrea and {Dayal}, Pratika},
        title = "{On the stunning abundance of super-early, luminous galaxies revealed by JWST}",
      journal = {\mnras},
         year = 2023,
        month = jul,
       volume = {522},
       number = {3},
        pages = {3986-3991},
          doi = {10.1093/mnras/stad1095},
archivePrefix = {arXiv},
       eprint = {2208.00720},
 primaryClass = {astro-ph.GA},
       adsurl = {https://ui.adsabs.harvard.edu/abs/2023MNRAS.522.3986F}
}

@ARTICLE{Ferrara25a,
       author = {{Ferrara}, A. and {Carniani}, S. and {di Mascia}, F. and {Bouwens}, R.~J. and {Oesch}, P. and {Schouws}, S.},
        title = "{ALMA observations of super-early galaxies: Attenuation-free model predictions}",
      journal = {\aap},
         year = 2025,
        month = feb,
       volume = {694},
          eid = {A215},
        pages = {A215},
          doi = {10.1051/0004-6361/202452368},
archivePrefix = {arXiv},
       eprint = {2409.17223},
 primaryClass = {astro-ph.GA},
       adsurl = {https://ui.adsabs.harvard.edu/abs/2025A&A...694A.215F}
}

@ARTICLE{Ferrara25b,
       author = {{Ferrara}, A. and {Pallottini}, A. and {Sommovigo}, L.},
        title = "{Blue monsters at z > 10: Where all their dust has gone}",
      journal = {\aap},
         year = 2025,
        month = feb,
       volume = {694},
          eid = {A286},
        pages = {A286},
          doi = {10.1051/0004-6361/202452707},
archivePrefix = {arXiv},
       eprint = {2410.19042},
 primaryClass = {astro-ph.GA},
       adsurl = {https://ui.adsabs.harvard.edu/abs/2025A&A...694A.286F}
}

@ARTICLE{Ferrara26,
       author = {{Ferrara}, A. and {Rodighiero}, G. and {Carniani}, S. and {Zhang}, Z. and {Kohandel}, M. and {Das}, B.},
        title = "{No Blue without Red: Evolutionary Properties of Super-Early Galaxies}",
      journal = {arXiv e-prints},
         year = 2026,
        month = may,
          eid = {arXiv:2605.22914},
        pages = {arXiv:2605.22914},
          doi = {10.48550/arXiv.2605.22914},
archivePrefix = {arXiv},
       eprint = {2605.22914},
 primaryClass = {astro-ph.GA},
       adsurl = {https://ui.adsabs.harvard.edu/abs/2026arXiv260522914F}
}

@ARTICLE{Finkelstein24,
       author = {{Finkelstein}, Steven L. and {Leung}, Gene C.~K. and {Bagley}, Micaela B. and {Dickinson}, Mark and {Ferguson}, Henry C. and {Papovich}, Casey and {Akins}, Hollis B. and {Arrabal Haro}, Pablo and {Dav{\'e}}, Romeel and {Dekel}, Avishai and {Kartaltepe}, Jeyhan S. and {Kocevski}, Dale D. and {Koekemoer}, Anton M. and {Pirzkal}, Nor and {Somerville}, Rachel S. and {Yung}, L.~Y. Aaron and {Amor{\'\i}n}, Ricardo O. and {Backhaus}, Bren E. and {Behroozi}, Peter and {Bisigello}, Laura and {Bromm}, Volker and {Casey}, Caitlin M. and {Ch{\'a}vez Ortiz}, {\'O}scar A. and {Cheng}, Yingjie and {Chworowsky}, Katherine and {Cleri}, Nikko J. and {Cooper}, M.~C. and {Davis}, Kelcey and {de la Vega}, Alexander and {Elbaz}, David and {Franco}, Maximilien and {Fontana}, Adriano and {Fujimoto}, Seiji and {Giavalisco}, Mauro and {Grogin}, Norman A. and {Holwerda}, Benne W. and {Huertas-Company}, Marc and {Hirschmann}, Michaela and {Iyer}, Kartheik G. and {Jogee}, Shardha and {Jung}, Intae and {Larson}, Rebecca L. and {Lucas}, Ray A. and {Mobasher}, Bahram and {Morales}, Alexa M. and {Morley}, Caroline V. and {Mukherjee}, Sagnick and {P{\'e}rez-Gonz{\'a}lez}, Pablo G. and {Ravindranath}, Swara and {Rodighiero}, Giulia and {Rowland}, Melanie J. and {Tacchella}, Sandro and {Taylor}, Anthony J. and {Trump}, Jonathan R. and {Wilkins}, Stephen M.},
        title = "{The Complete CEERS Early Universe Galaxy Sample: A Surprisingly Slow Evolution of the Space Density of Bright Galaxies at z {\ensuremath{\sim}} 8.5{\textendash}14.5}",
      journal = {\apjl},
         year = 2024,
        month = jul,
       volume = {969},
       number = {1},
          eid = {L2},
        pages = {L2},
          doi = {10.3847/2041-8213/ad4495},
archivePrefix = {arXiv},
       eprint = {2311.04279},
 primaryClass = {astro-ph.GA},
       adsurl = {https://ui.adsabs.harvard.edu/abs/2024ApJ...969L...2F}
}

@ARTICLE{fontana2026,
       author = {{Fontana}, A. and {Atek}, H. and {Bromm}, V. and {Bunker}, A.~J. and {Castellano}, M. and {Dayal}, P. and {D'Eugenio}, F. and {Elbaz}, D. and {Ferrara}, A. and {Finkelstein}, S.~L. and {Fujimoto}, S. and {Giavalisco}, M. and {Harikane}, Y. and {Kobayashi}, C. and {Pacucci}, F. and {P{\'e}rez-Gonz{\'a}lez}, P.~G. and {Rosati}, P. and {Trinca}, A. and {Tripodi}, R. and {Santini}, P. and {Vanzella}, E. and {Venditti}, A. and {Weibel}, A.},
        title = "{Too many or too bright? Status and perspectives in the study of the UV Luminosity Function at z>10}",
      journal = {arXiv e-prints},
         year = 2026,
        month = oct,
          eid = {arXiv:2610.12121},
        pages = {arXiv:2610.12121},
          doi = {10.48550/arXiv.2610.12121},
archivePrefix = {arXiv},
       eprint = {2610.12121},
 primaryClass = {astro-ph.GA},
       adsurl = {https://ui.adsabs.harvard.edu/abs/2026arXiv261012121F}
}

@ARTICLE{Harikane22,
       author = {{Harikane}, Yuichi and {Ono}, Yoshiaki and {Ouchi}, Masami and {Liu}, Chengze and {Sawicki}, Marcin and {Shibuya}, Takatoshi and {Behroozi}, Peter S. and {He}, Wanqiu and {Shimasaku}, Kazuhiro and {Arnouts}, Stephane and {Coupon}, Jean and {Fujimoto}, Seiji and {Gwyn}, Stephen and {Huang}, Jiasheng and {Inoue}, Akio K. and {Kashikawa}, Nobunari and {Komiyama}, Yutaka and {Matsuoka}, Yoshiki and {Willott}, Chris J.},
        title = "{GOLDRUSH. IV. Luminosity Functions and Clustering Revealed with  4,000,000 Galaxies at z   2-7: Galaxy-AGN Transition, Star Formation Efficiency, and Implication for Evolution at z > 10}",
      journal = {\apjs},
         year = 2022,
        month = mar,
       volume = {259},
       number = {1},
          eid = {20},
        pages = {20},
          doi = {10.3847/1538-4365/ac3dfc},
archivePrefix = {arXiv},
       eprint = {2108.01090},
 primaryClass = {astro-ph.GA},
       adsurl = {https://ui.adsabs.harvard.edu/abs/2022ApJS..259...20H}
}

@ARTICLE{Harikane24,
       author = {{Harikane}, Yuichi and {Nakajima}, Kimihiko and {Ouchi}, Masami and {Umeda}, Hiroya and {Isobe}, Yuki and {Ono}, Yoshiaki and {Xu}, Yi and {Zhang}, Yechi},
        title = "{Pure Spectroscopic Constraints on UV Luminosity Functions and Cosmic Star Formation History from 25 Galaxies at z $_{spec}$ = 8.61-13.20 Confirmed with JWST/NIRSpec}",
      journal = {\apj},
         year = 2024,
        month = jan,
       volume = {960},
       number = {1},
          eid = {56},
        pages = {56},
          doi = {10.3847/1538-4357/ad0b7e},
archivePrefix = {arXiv},
       eprint = {2304.06658},
 primaryClass = {astro-ph.GA},
       adsurl = {https://ui.adsabs.harvard.edu/abs/2024ApJ...960...56H}
}

@ARTICLE{Harikane26,
       author = {{Harikane}, Yuichi and {Perez-Gonzalez}, Pablo G. and {Alvarez-Marquez}, Javier and {Ouchi}, Masami and {Nakazato}, Yurina and {Ono}, Yoshiaki and {Nakajima}, Kimihiko and {Umeda}, Hiroya and {Isobe}, Yuki and {Xu}, Yi and {Zhang}, Yechi},
        title = "{A UV-Luminous Galaxy at z=11 with Surprisingly Weak Star Formation Activity}",
      journal = {arXiv e-prints},
         year = 2026,
        month = jan,
          eid = {arXiv:2601.21833},
        pages = {arXiv:2601.21833},
          doi = {10.48550/arXiv.2601.21833},
archivePrefix = {arXiv},
       eprint = {2601.21833},
 primaryClass = {astro-ph.GA},
       adsurl = {https://ui.adsabs.harvard.edu/abs/2026arXiv260121833H}
}

@ARTICLE{Helton25,
        author = {{Helton}, Jakob M. and {Rieke}, George H. and {Alberts}, Stacey and {Wu}, Zihao and {Eisenstein}, Daniel J. and {Hainline}, Kevin N. and {Carniani}, Stefano and {Ji}, Zhiyuan and {Baker}, William M. and {Bhatawdekar}, Rachana and {Bunker}, Andrew J. and {Cargile}, Phillip A. and {Charlot}, St{\'e}phane and {Chevallard}, Jacopo and {D'Eugenio}, Francesco and {Egami}, Eiichi and {Johnson}, Benjamin D. and {Jones}, Gareth C. and {Lyu}, Jianwei and {Maiolino}, Roberto and {P{\'e}rez-Gonz{\'a}lez}, Pablo G. and {Rieke}, Marcia J. and {Robertson}, Brant and {Saxena}, Aayush and {Scholtz}, Jan and {Shivaei}, Irene and {Sun}, Fengwu and {Tacchella}, Sandro and {Whitler}, Lily and {Williams}, Christina C. and {Willmer}, Christopher N.~A. and {Willott}, Chris and {Witstok}, Joris and {Zhu}, Yongda},
        title = "{Photometric detection at 7.7 {\ensuremath{\mu}}m of a galaxy beyond redshift 14 with JWST/MIRI}",
      journal = {Nature Astronomy},
         year = 2025,
        month = may,
       volume = {9},
        pages = {729-740},
          doi = {10.1038/s41550-025-02503-z},
archivePrefix = {arXiv},
       eprint = {2405.18462},
 primaryClass = {astro-ph.GA},
       adsurl = {https://ui.adsabs.harvard.edu/abs/2025NatAs...9..729H}
}

@article{Hennebelle13,
  author  = {Hennebelle, Patrick and Chabrier, Gilles},
  title   = {Analytical Theory for the Initial Mass Function. III. Time Dependence and Star Formation Rate},
  journal = {The Astrophysical Journal},
  year    = {2013},
  volume  = {770},
  pages   = {150},
  doi     = {10.1088/0004-637X/770/2/150},
  eprint  = {1304.6637},
  archivePrefix = {arXiv},
  primaryClass = {astro-ph.GA}
}

@ARTICLE{Hirano14,
       author = {{Hirano}, Shingo and {Hosokawa}, Takashi and {Yoshida}, Naoki and {Umeda}, Hideyuki and {Omukai}, Kazuyuki and {Chiaki}, Gen and {Yorke}, Harold W.},
        title = "{One Hundred First Stars: Protostellar Evolution and the Final Masses}",
      journal = {\apj},
         year = 2014,
        month = feb,
       volume = {781},
       number = {2},
          eid = {60},
        pages = {60},
          doi = {10.1088/0004-637X/781/2/60},
archivePrefix = {arXiv},
       eprint = {1308.4456},
 primaryClass = {astro-ph.CO},
       adsurl = {https://ui.adsabs.harvard.edu/abs/2014ApJ...781...60H}
}

@ARTICLE{Hirano15,
       author = {{Hirano}, S. and {Hosokawa}, T. and {Yoshida}, N. and {Omukai}, K. and {Yorke}, H.~W.},
        title = "{Primordial star formation under the influence of far ultraviolet radiation: 1540 cosmological haloes and the stellar mass distribution}",
      journal = {\mnras},
         year = 2015,
        month = mar,
       volume = {448},
       number = {1},
        pages = {568-587},
          doi = {10.1093/mnras/stv044},
archivePrefix = {arXiv},
       eprint = {1501.01630},
 primaryClass = {astro-ph.GA},
       adsurl = {https://ui.adsabs.harvard.edu/abs/2015MNRAS.448..568H}
}

@ARTICLE{Hirano26,
       author = {{Hirano}, Shingo and {Sakai}, Yusuke and {Umeda}, Hideyuki},
        title = "{Formation of first star clusters under the supersonic gas flow ─ III. Environmental trends and halo-to-halo scatter in the Pop III IMF}",
      journal = {\mnras},
         year = 2026,
        month = may,
       volume = {548},
       number = {3},
          doi = {10.1093/mnras/stag634},
archivePrefix = {arXiv},
       eprint = {2602.20690},
 primaryClass = {astro-ph.GA},
       adsurl = {https://ui.adsabs.harvard.edu/abs/2026MNRAS.548ag634H}
}

@ARTICLE{Hsiao24,
       author = {{Hsiao}, Tiger Yu-Yang and {{\'A}lvarez-M{\'a}rquez}, Javier and {Coe}, Dan and {Crespo G{\'o}mez}, Alejandro and {Abdurro'uf} and {Dayal}, Pratika and {Larson}, Rebecca L. and {Bik}, Arjan and {Blanco-Prieto}, Carmen and {Colina}, Luis and {P{\'e}rez-Gonz{\'a}lez}, Pablo Guillermo and {Costantin}, Luca and {Prieto-Jim{\'e}nez}, Carlota and {Adamo}, Angela and {Bradley}, Larry D. and {Conselice}, Christopher J. and {Fujimoto}, Seiji and {Furtak}, Lukas J. and {Hutchison}, Taylor A. and {James}, Bethan L. and {Jim{\'e}nez-Teja}, Yolanda and {Jung}, Intae and {Kokorev}, Vasily and {Mingozzi}, Matilde and {Norman}, Colin and {Ricotti}, Massimo and {Rigby}, Jane R. and {Sharon}, Keren and {Vanzella}, Eros and {Welch}, Brian and {Xu}, Xinfeng and {Zackrisson}, Erik and {Zitrin}, Adi},
        title = "{JWST MIRI Detections of H{\ensuremath{\alpha}} and [O III] and a Direct Metallicity Measurement of the z = 10.17 Lensed Galaxy MACS0647{\ensuremath{-}}JD}",
      journal = {\apj},
         year = 2024,
        month = oct,
       volume = {973},
       number = {2},
          eid = {81},
        pages = {81},
          doi = {10.3847/1538-4357/ad6562},
archivePrefix = {arXiv},
       eprint = {2404.16200},
 primaryClass = {astro-ph.GA},
       adsurl = {https://ui.adsabs.harvard.edu/abs/2024ApJ...973...81H}
}

@ARTICLE{Hutter_2024,
       author = {{Hutter}, Anne and {Cueto}, Elie R. and {Dayal}, Pratika and {Gottl{\"o}ber}, Stefan and {Trebitsch}, Maxime and {Yepes}, Gustavo},
        title = "{ASTRAEUS: X. Indications of a top-heavy initial mass function in highly star-forming galaxies from JWST observations at z > 10}",
      journal = {\aap},
         year = 2025,
        month = feb,
       volume = {694},
          eid = {A254},
        pages = {A254},
          doi = {10.1051/0004-6361/202452460},
archivePrefix = {arXiv},
       eprint = {2410.00730},
 primaryClass = {astro-ph.GA},
       adsurl = {https://ui.adsabs.harvard.edu/abs/2025A&A...694A.254H}
}

@article{Jerabkova18,
  author  = {Je{\v{r}}{\'a}bkov{\'a}, Tereza and Kroupa, Pavel and Dabringhausen, J{\"o}rg and Hilker, Michael and Bekki, Kenji},
  title   = {The impact of the metallicity and star formation rate on the time-dependent galaxy-wide stellar initial mass function},
  journal = {Astronomy \& Astrophysics},
  year    = {2018},
  volume  = {620},
  pages   = {A39},
  doi     = {10.1051/0004-6361/201833055},
  eprint  = {1809.04603},
  archivePrefix = {arXiv},
  primaryClass = {astro-ph.GA}
}

@ARTICLE{Katz23,
       author = {{Katz}, Harley and {Kimm}, Taysun and {Ellis}, Richard S. and {Devriendt}, Julien and {Slyz}, Adrianne},
        title = "{The challenges of identifying Population III stars in the early Universe}",
      journal = {\mnras},
         year = 2023,
        month = sep,
       volume = {524},
       number = {1},
        pages = {351-360},
          doi = {10.1093/mnras/stad1903},
archivePrefix = {arXiv},
       eprint = {2207.04751},
 primaryClass = {astro-ph.GA},
       adsurl = {https://ui.adsabs.harvard.edu/abs/2023MNRAS.524..351K}
}

@ARTICLE{Katz25,
       author = {{Katz}, Harley and {Cameron}, Alex J. and {Saxena}, Aayush and {Barrufet}, Laia and {Choustikov}, Nichloas and {Cleri}, Nikko J. and {de Graff}, Anna and {Ellis}, Richard S. and {Fosbury}, Robert A.~E. and {Heintz}, Kasper E. and {Maseda}, Michael and {Matthee}, Jorryt and {McConachie}, Ian and {Oesch}, Pascal A.},
        title = "{21 Balmer Jump Street: The Nebular Continuum at High Redshift and Implications for the Bright Galaxy Problem, UV Continuum Slopes, and Early Stellar Populations}",
      journal = {The Open Journal of Astrophysics},
         year = 2025,
        month = jul,
       volume = {8},
          eid = {104},
        pages = {104},
          doi = {10.33232/001c.142570},
archivePrefix = {arXiv},
       eprint = {2408.03189},
 primaryClass = {astro-ph.GA},
       adsurl = {https://ui.adsabs.harvard.edu/abs/2025OJAp....8E.104K}
}

@ARTICLE{Kim14,
       author = {{Kim}, Ji-hoon and {Abel}, Tom and {Agertz}, Oscar and {Bryan}, Greg L. and {Ceverino}, Daniel and {Christensen}, Charlotte and {Conroy}, Charlie and {Dekel}, Avishai and {Gnedin}, Nickolay Y. and {Goldbaum}, Nathan J. and {Guedes}, Javiera and {Hahn}, Oliver and {Hobbs}, Alexander and {Hopkins}, Philip F. and {Hummels}, Cameron B. and {Iannuzzi}, Francesca and {Keres}, Dusan and {Klypin}, Anatoly and {Kravtsov}, Andrey V. and {Krumholz}, Mark R. and {Kuhlen}, Michael and {Leitner}, Samuel N. and {Madau}, Piero and {Mayer}, Lucio and {Moody}, Christopher E. and {Nagamine}, Kentaro and {Norman}, Michael L. and {Onorbe}, Jose and {O'Shea}, Brian W. and {Pillepich}, Annalisa and {Primack}, Joel R. and {Quinn}, Thomas and {Read}, Justin I. and {Robertson}, Brant E. and {Rocha}, Miguel and {Rudd}, Douglas H. and {Shen}, Sijing and {Smith}, Britton D. and {Szalay}, Alexander S. and {Teyssier}, Romain and {Thompson}, Robert and {Todoroki}, Keita and {Turk}, Matthew J. and {Wadsley}, James W. and {Wise}, John H. and {Zolotov}, Adi and {AGORA Collaboration29},the},
        title = "{The AGORA High-resolution Galaxy Simulations Comparison Project}",
      journal = {\apjs},
         year = 2014,
        month = jan,
       volume = {210},
       number = {1},
          eid = {14},
        pages = {14},
          doi = {10.1088/0067-0049/210/1/14},
archivePrefix = {arXiv},
       eprint = {1308.2669},
 primaryClass = {astro-ph.GA},
       adsurl = {https://ui.adsabs.harvard.edu/abs/2014ApJS..210...14K}
}

@ARTICLE{kreilgaard2026,
       author = {{Kreilgaard}, Kimi C. and {Mason}, Charlotte A. and {Morishita}, Takahiro and {Zhang}, Yechi and {Gelli}, Viola and {Leethochawalit}, Nicha and {Treu}, Tommaso and {Trenti}, Michele and {Abdurro'uf} and {Atek}, Hakim and {Bradac}, Maru{\v{s}}a and {Bradley}, Larry D. and {Bunker}, Andrew J. and {Haryana}, Novan Saputra and {Hayes}, Matthew J. and {Liu}, Zhaoran and {Mehta}, Vihang and {Rafelski}, Marc and {Roberts-Borsani}, Guido and {Scarlata}, Claudia and {Stiavelli}, Massimo and {Sutanto}, Ryo A. and {Takahashi}, Kosuke and {Vulcani}, Benedetta},
        title = "{BEACON: JWST NIRCam Pure-parallel Imaging Survey. III. Constraints on the Ultraviolet Luminosity Function and the Clustering of z \raisebox{-0.5ex}\textasciitilde 7-14 Galaxies}",
      journal = {\apj},
         year = 2026,
        month = oct,
       volume = {1009},
       number = {2},
          eid = {134},
        pages = {134},
          doi = {10.3847/1538-4357/ae9152},
archivePrefix = {arXiv},
       eprint = {2604.17963},
 primaryClass = {astro-ph.GA},
       adsurl = {https://ui.adsabs.harvard.edu/abs/2026ApJ..1009..134K}
}

@ARTICLE{Lesniewska19,
       author = {{Le{\'s}niewska}, Aleksandra and {Micha{\l}owski}, Micha{\l} Jerzy},
        title = "{Dust production scenarios in galaxies at z ̃6-8.3}",
      journal = {\aap},
         year = 2019,
        month = apr,
       volume = {624},
          eid = {L13},
        pages = {L13},
          doi = {10.1051/0004-6361/201935149},
archivePrefix = {arXiv},
       eprint = {1904.11185},
 primaryClass = {astro-ph.GA},
       adsurl = {https://ui.adsabs.harvard.edu/abs/2019A&A...624L..13L}
}

@ARTICLE{Leung23,
       author = {{Leung}, Gene C.~K. and {Bagley}, Micaela B. and {Finkelstein}, Steven L. and {Ferguson}, Henry C. and {Koekemoer}, Anton M. and {P{\'e}rez-Gonz{\'a}lez}, Pablo G. and {Morales}, Alexa and {Kocevski}, Dale D. and {Yang}, Guang and {Somerville}, Rachel S. and {Wilkins}, Stephen M. and {Yung}, L.~Y. Aaron and {Fujimoto}, Seiji and {Larson}, Rebecca L. and {Papovich}, Casey and {Pirzkal}, Nor and {Berg}, Danielle A. and {Lotz}, Jennifer M. and {Castellano}, Marco and {Ch{\'a}vez Ortiz}, {\'O}scar A. and {Cheng}, Yingjie and {Dickinson}, Mark and {Giavalisco}, Mauro and {Hathi}, Nimish P. and {Hutchison}, Taylor A. and {Jung}, Intae and {Kartaltepe}, Jeyhan S. and {Natarajan}, Priyamvada and {Rothberg}, Barry},
        title = "{NGDEEP Epoch 1: The Faint End of the Luminosity Function at z   9-12 from Ultradeep JWST Imaging}",
      journal = {\apjl},
         year = 2023,
        month = sep,
       volume = {954},
       number = {2},
          eid = {L46},
        pages = {L46},
          doi = {10.3847/2041-8213/acf365},
archivePrefix = {arXiv},
       eprint = {2306.06244},
 primaryClass = {astro-ph.GA},
       adsurl = {https://ui.adsabs.harvard.edu/abs/2023ApJ...954L..46L}
}

@ARTICLE{Li24,
       author = {{Li}, Yijia and {Leja}, Joel and {Johnson}, Benjamin D. and {Tacchella}, Sandro and {Naidu}, Rohan P.},
        title = "{No Top-heavy Stellar Initial Mass Function Needed: The Ionizing Radiation of GS9422 Can Be Powered by a Mixture of an Active Galactic Nucleus and Stars}",
      journal = {\apjl},
         year = 2024,
        month = jul,
       volume = {969},
       number = {1},
          eid = {L5},
        pages = {L5},
          doi = {10.3847/2041-8213/ad5280},
archivePrefix = {arXiv},
       eprint = {2404.02333},
 primaryClass = {astro-ph.GA},
       adsurl = {https://ui.adsabs.harvard.edu/abs/2024ApJ...969L...5L}
}

@article{Marks12,
  author  = {Marks, Michael and Kroupa, Pavel and Dabringhausen, J{\"o}rg and Pawlowski, Marcel S.},
  title   = {Evidence for top-heavy stellar initial mass functions with increasing density and decreasing metallicity},
  journal = {Monthly Notices of the Royal Astronomical Society},
  year    = {2012},
  volume  = {422},
  pages   = {2246--2254},
  doi     = {10.1111/j.1365-2966.2012.20767.x},
  eprint  = {1202.4755},
  archivePrefix = {arXiv},
  primaryClass = {astro-ph.GA}
}

@ARTICLE{Mathew26,
       author = {{Mathew}, Sajay Sunny and {Federrath}, Christoph and {Seta}, Amit},
        title = "{The impact of the turbulent Mach number on star formation and the initial mass function}",
      journal = {arXiv e-prints},
         year = 2026,
        month = sep,
          eid = {arXiv:2609.38937},
        pages = {arXiv:2609.38937},
          doi = {10.48550/arXiv.2609.38937},
archivePrefix = {arXiv},
       eprint = {2609.38937},
 primaryClass = {astro-ph.GA},
       adsurl = {https://ui.adsabs.harvard.edu/abs/2026arXiv260938937M}
}

@ARTICLE{mauerhofer2025,
       author = {{Mauerhofer}, Valentin and {Dayal}, Pratika and {Haehnelt}, Martin G. and {Kimm}, Taysun and {Rosdahl}, Joakim and {Teyssier}, Romain},
        title = "{Synergising semi-analytical models and hydrodynamical simulations to interpret JWST data from the first billion years}",
      journal = {\aap},
         year = 2025,
        month = apr,
       volume = {696},
          eid = {A157},
        pages = {A157},
          doi = {10.1051/0004-6361/202554042},
archivePrefix = {arXiv},
       eprint = {2502.02647},
 primaryClass = {astro-ph.GA},
       adsurl = {https://ui.adsabs.harvard.edu/abs/2025A&A...696A.157M}
}

@ARTICLE{Mcleod24,
       author = {{McLeod}, D.~J. and {Donnan}, C.~T. and {McLure}, R.~J. and {Dunlop}, J.~S. and {Magee}, D. and {Begley}, R. and {Carnall}, A.~C. and {Cullen}, F. and {Ellis}, R.~S. and {Hamadouche}, M.~L. and {Stanton}, T.~M.},
        title = "{The galaxy UV luminosity function at z ≃ 11 from a suite of public JWST ERS, ERO, and Cycle-1 programs}",
      journal = {\mnras},
         year = 2024,
        month = jan,
       volume = {527},
       number = {3},
        pages = {5004-5022},
          doi = {10.1093/mnras/stad3471},
archivePrefix = {arXiv},
       eprint = {2304.14469},
 primaryClass = {astro-ph.GA},
       adsurl = {https://ui.adsabs.harvard.edu/abs/2024MNRAS.527.5004M}
}

@ARTICLE{Menon24,
       author = {{Menon}, Shyam H. and {Lancaster}, Lachlan and {Burkhart}, Blakesley and {Somerville}, Rachel S. and {Dekel}, Avishai and {Krumholz}, Mark R.},
        title = "{The Interplay between the Initial Mass Function and Star Formation Efficiency through Radiative Feedback at High Stellar Surface Densities}",
      journal = {\apjl},
         year = 2024,
        month = jun,
       volume = {967},
       number = {2},
          eid = {L28},
        pages = {L28},
          doi = {10.3847/2041-8213/ad462d},
archivePrefix = {arXiv},
       eprint = {2405.00813},
 primaryClass = {astro-ph.GA},
       adsurl = {https://ui.adsabs.harvard.edu/abs/2024ApJ...967L..28M}
}

@ARTICLE{Naidu22,
       author = {{Naidu}, Rohan P. and {Oesch}, Pascal A. and {van Dokkum}, Pieter and {Nelson}, Erica J. and {Suess}, Katherine A. and {Brammer}, Gabriel and {Whitaker}, Katherine E. and {Illingworth}, Garth and {Bouwens}, Rychard and {Tacchella}, Sandro and {Matthee}, Jorryt and {Allen}, Natalie and {Bezanson}, Rachel and {Conroy}, Charlie and {Labbe}, Ivo and {Leja}, Joel and {Leonova}, Ecaterina and {Magee}, Dan and {Price}, Sedona H. and {Setton}, David J. and {Strait}, Victoria and {Stefanon}, Mauro and {Toft}, Sune and {Weaver}, John R. and {Weibel}, Andrea},
        title = "{Two Remarkably Luminous Galaxy Candidates at z {\ensuremath{\approx}} 10-12 Revealed by JWST}",
      journal = {\apjl},
         year = 2022,
        month = nov,
       volume = {940},
       number = {1},
          eid = {L14},
        pages = {L14},
          doi = {10.3847/2041-8213/ac9b22},
archivePrefix = {arXiv},
       eprint = {2207.09434},
 primaryClass = {astro-ph.GA},
       adsurl = {https://ui.adsabs.harvard.edu/abs/2022ApJ...940L..14N}
}

@ARTICLE{Naidu25,
       author = {{Naidu}, Rohan P. and {Oesch}, Pascal A. and {Brammer}, Gabriel and {Weibel}, Andrea and {Li}, Yijia and {Matthee}, Jorryt and {Chisholm}, John and {Pollock}, Clara L. and {Heintz}, Kasper E. and {Johnson}, Benjamin D. and {Shen}, Xuejian and {Hviding}, Raphael E. and {Leja}, Joel and {Tacchella}, Sandro and {Ganguly}, Arpita and {Witten}, Callum and {Atek}, Hakim and {Belli}, Sirio and {Bose}, Sownak and {Bouwens}, Rychard and {Dayal}, Pratika and {Decarli}, Roberto and {de Graaff}, Anna and {Fudamoto}, Yoshinobu and {Giovinazzo}, Emma and {Greene}, Jenny E. and {Illingworth}, Garth and {Inoue}, Akio K. and {Kane}, Sarah G. and {Labbe}, Ivo and {Leonova}, Ecaterina and {Marques-Chaves}, Rui and {Meyer}, Romain A. and {Nelson}, Erica J. and {Roberts-Borsani}, Guido and {Schaerer}, Daniel and {Simcoe}, Robert A. and {Stefanon}, Mauro and {Sugahara}, Yuma and {Toft}, Sune and {van der Wel}, Arjen and {van Dokkum}, Pieter and {Walter}, Fabian and {Watson}, Darach and {Weaver}, John R. and {Whitaker}, Katherine E.},
        title = "{A Cosmic Miracle: A Remarkably Luminous Galaxy at $z_{\rm{spec}}=14.44$ Confirmed with JWST}",
      journal = {arXiv e-prints},
         year = 2025,
        month = may,
          eid = {arXiv:2505.11263},
        pages = {arXiv:2505.11263},
          doi = {10.48550/arXiv.2505.11263},
archivePrefix = {arXiv},
       eprint = {2505.11263},
 primaryClass = {astro-ph.GA},
       adsurl = {https://ui.adsabs.harvard.edu/abs/2025arXiv250511263N}
}

@article{Narayanan12,
  author  = {Narayanan, Desika and Dav{\'e}, Romeel},
  title   = {Cosmological implications of a stellar initial mass function that varies with the Jeans mass in galaxies},
  journal = {Monthly Notices of the Royal Astronomical Society},
  year    = {2012},
  volume  = {423},
  pages   = {3601--3612},
  doi     = {10.1111/j.1365-2966.2012.21119.x},
  eprint  = {1203.1039},
  archivePrefix = {arXiv},
  primaryClass = {astro-ph.CO}
}

@ARTICLE{Nomoto13,
       author = {{Nomoto}, Ken'ichi and {Kobayashi}, Chiaki and {Tominaga}, Nozomu},
        title = "{Nucleosynthesis in Stars and the Chemical Enrichment of Galaxies}",
      journal = {\araa},
         year = 2013,
        month = aug,
       volume = {51},
       number = {1},
        pages = {457-509},
          doi = {10.1146/annurev-astro-082812-140956},
       adsurl = {https://ui.adsabs.harvard.edu/abs/2013ARA&A..51..457N}
}

@ARTICLE{Omukai05,
       author = {{Omukai}, K. and {Tsuribe}, T. and {Schneider}, R. and {Ferrara}, A.},
        title = "{Thermal and Fragmentation Properties of Star-forming Clouds in Low-Metallicity Environments}",
      journal = {\apj},
         year = 2005,
        month = jun,
       volume = {626},
       number = {2},
        pages = {627-643},
          doi = {10.1086/429955},
archivePrefix = {arXiv},
       eprint = {astro-ph/0503010},
 primaryClass = {astro-ph},
       adsurl = {https://ui.adsabs.harvard.edu/abs/2005ApJ...626..627O}
}

@ARTICLE{Omukai98,
       author = {{Omukai}, Kazuyuki and {Nishi}, Ryoichi},
        title = "{Formation of Primordial Protostars}",
      journal = {\apj},
         year = 1998,
        month = nov,
       volume = {508},
       number = {1},
        pages = {141-150},
          doi = {10.1086/306395},
archivePrefix = {arXiv},
       eprint = {astro-ph/9811308},
 primaryClass = {astro-ph},
       adsurl = {https://ui.adsabs.harvard.edu/abs/1998ApJ...508..141O}
}

@article{Padoan97,
  author  = {Padoan, Paolo and Nordlund, {\AA}ke and Jones, Bernard J. T.},
  title   = {The universality of the stellar initial mass function},
  journal = {Monthly Notices of the Royal Astronomical Society},
  year    = {1997},
  volume  = {288},
  pages   = {145--152},
  doi     = {10.1093/mnras/288.1.145}
}


\end{document}